\documentclass[utf8]{FrontiersinHarvard} 
\usepackage{url,hyperref,lineno}
\usepackage[onehalfspacing]{setspace}
\usepackage{datetime}
\usepackage{fmtcount}
\usepackage{etoolbox}
\usepackage{fcprefix}
\usepackage{graphicx}
\usepackage{textcomp, gensymb}
\usepackage{siunitx}
\usepackage{natbib}
\usepackage{romannum}

\def\keyFont{\fontsize{8}{11}\helveticabold }
\def\firstAuthorLast{Mu {et~al.}} 
\def\Authors{Xin Mu\,$^{1,2,*}$, Fu-Der Chen\,$^{1,2,3}$, Ka My Dang\,$^{1,3}$, Michael G. K. Brunk\,$^{1,3}$, Jianfeng Li\,$^{1,3}$, Hannes Wahn\,$^{1}$, Andrei Stalmashonak\,$^{1}$, Peisheng Ding\,$^{1,2}$, Xianshu Luo\,$^{4}$, Hongyao Chua\,$^{4}$, Guo-Qiang Lo\,$^{4}$, Joyce K. S. Poon\,$^{1,2,3}$ and Wesley D. Sacher\,$^{1,3,*}$}

\def\Address{$^{1}$Max Planck Institute of Microstructure Physics, Halle, Germany \\
$^{2}$Department of Electrical and Computer Engineering, University of Toronto, Toronto, ON, Canada\\
$^{3}$Max Planck-University of Toronto Centre for Neural Science and Technology\\
$^{4}$Advanced Micro Foundry Pte. Ltd., Singapore, Singapore}

\def\corrAuthor{Xin Mu}
\def\corrEmail{xinmu@mpi-halle.mpg.de}

\begin{document}
\onecolumn
\firstpage{1}
\pagenumbering{arabic}

\title {Implantable Photonic Neural Probes with 3D-Printed Microfluidics and Applications to Uncaging}

\author[\firstAuthorLast ]{\Authors} 
\address{} 
\correspondance{} 
\extraAuth{Wesley D. Sacher \\wesley.sacher@mpi-halle.mpg.de}

\graphicspath{{figures/}}
\maketitle
\begin{abstract}
\section{}

Advances in chip-scale photonic-electronic integration are enabling a new generation of foundry-manufacturable implantable silicon neural probes incorporating nanophotonic waveguides and microelectrodes for optogenetic stimulation and electrophysiological recording in neuroscience research. Further extending neural probe functionalities with integrated microfluidics is a direct approach to achieve neurochemical injection and sampling capabilities. In this work, we use two-photon polymerization 3D printing to integrate microfluidic channels onto photonic neural probes, which include silicon nitride nanophotonic waveguides and grating emitters. The customizability of 3D printing enables a unique geometry of microfluidics that conforms to the shape of each neural probe, enabling integration of microfluidics with a variety of existing neural probes while avoiding the complexities of monolithic microfluidics integration. We demonstrate the photonic and fluidic functionalities of the neural probes via fluorescein injection in agarose gel and photoloysis of caged fluorescein in solution and in flxed brain tissue. 

\tiny
 \keyFont{ Keywords: neural probe, neural interface, implantable device, 3D printing, microfluidics, laser photolysis, uncaging} 
\end{abstract}

\section{Introduction}

In recent years, a variety of implantable silicon (Si) neural probes have been developed with dense arrays of microelectrodes for electrophysiological recordings \citep{Rios_2016, li2018nanofabricated, steinmetz2021neuropixels}, nanophotonic waveguides for optogenetic stimulation \citep{Shim_2016, Segev_2016, sacher2019visible}, and co-integrated waveguides (or micro light-emitting diodes) and microelectrodes \citep{wu2015monolithically, Libbrecht_2018, Mohanty_2020, chen2022implantable}. These neural probes are capable of high spatiotemporal resolution optogenetic stimulation and/or extracellular recordings at depth within the brain while displacing relatively small volumes of tissue upon implantation - enabling new neuroscience experiments free of the limitations in depth, resolution, and/or tissue displacement imposed by conventional tools, e.g., implanted optical fibers \citep{Pisanello_2018}, gradient refractive index (GRIN) lenses \citep{Accanto_2019}, miniaturized microscopes \citep{Aharoni_2019}, and tetrodes \citep{Voigts_2013}. Parallel efforts have developed neural probes with integrated microfluidic channels for fluid injection and sampling of extracellular fluids \citep{Lee_2013,lee2015multichannel,shin2019multifunctional}, importantly, providing direct access to the neurochemical domain. Of particular interest are neuropharmacological applications wherein pharmacological agents are delivered to deep brain regions, for studies of complex neural circuits and neurotransmitter/receptor reaction mechanisms toward the treatment of neurodegenerative diseases \citep{sim2017microfluidic}. Previous demonstrations of microfluidic neural probes included either no \citep{Lee_2013} or relatively limited photonic/electrophysiological functionalities (32 electrodes and 1 large polymer waveguide) \citep{shin2019multifunctional}. Overall, integrating microfluidics onto neural probes with state of the art photonic and electrophysiological capabilities has remained an open challenge.

\begin{figure}[ht]
\begin{center}
\includegraphics[width=1\textwidth]{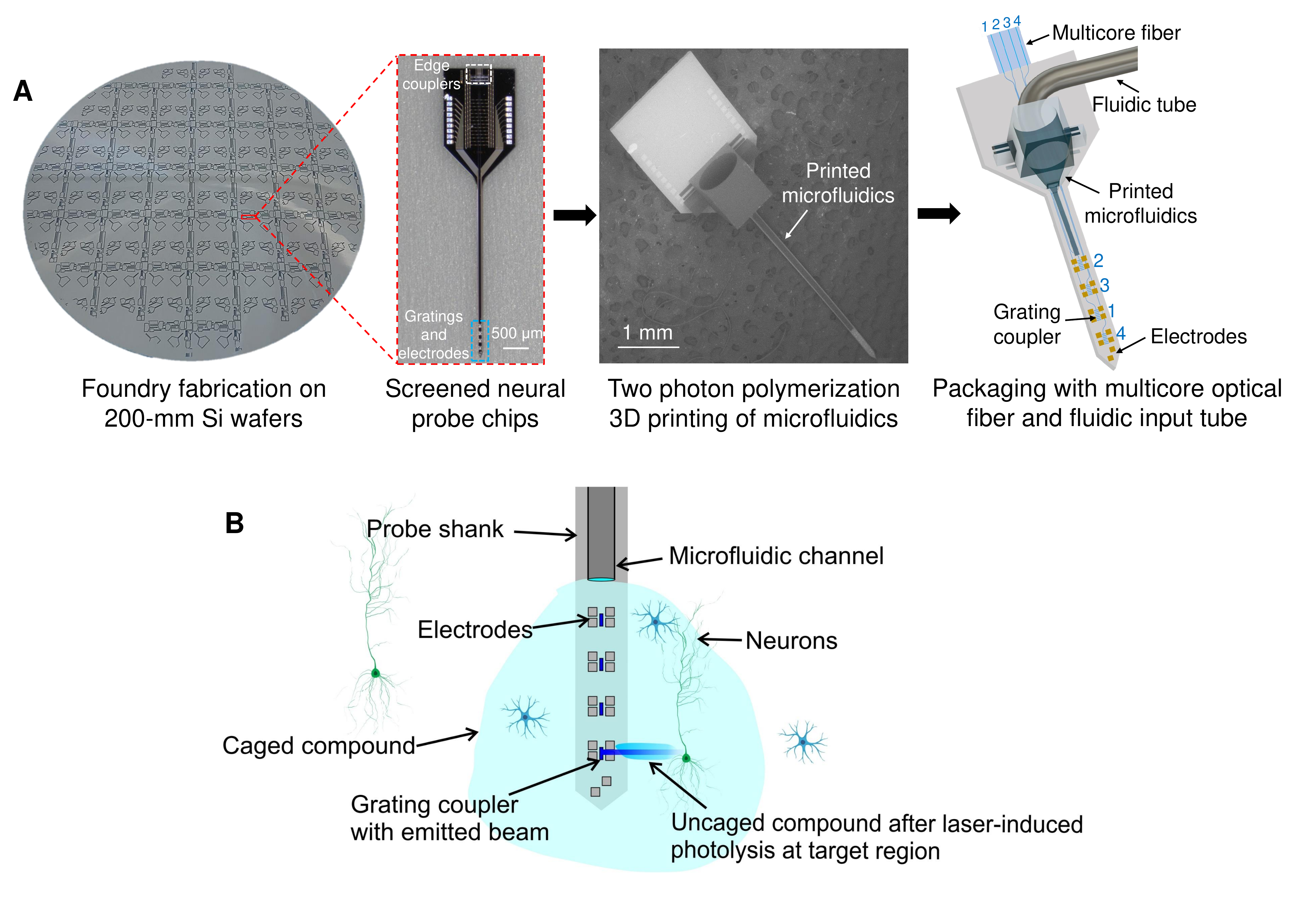}%
\end{center}
\caption{\textbf{(A)} Our approach for integration of microfluidic channels with foundry-fabricated photonic neural probes. Neural probe chips from 200-mm diameter wafers are screened, then two photon polymerization 3D printing of microfluidic channels and connectors is performed on each probe chip, and finally, the probe chips are packaged with multicore optical fibers and fluidic tubes. \textbf{(Left)} Photograph of a fabricated neural probe wafer (photograph background was removed for visibility), \textbf{(center-left)} optical micrograph of a neural probe chip, \textbf{(center-right)} scanning electron micrograph (SEM) of a neural probe chip with a 3D-printed microfluidic channel and connector, \textbf{(right)} illustration of a packaged neural probe with an input multicore fiber and fluidic tube (not to scale). \textbf{(B)} Conceptual illustration of simultaneous fluid injection, extracellular recording, and optogenetic stimulation (not to scale). Local photolysis of a caged compound injected by the microfluidic channel and uncaged by the addressable laser beams emitted by the probe is shown.}
\label{fig:1}
\end{figure}

In this work, we present a method of integrating microfluidic structures onto neural probes using additive two photon polymerization 3D printing \citep{zhang2020ultra, luo2021biologically, sun2022assembly}. This 3D printing technology allows for a high degree of flexibility in the manufacturing process \citep{lolsberg20183d}, and the layer-by-layer construction of 3D microstructures (through photocurable material crosslinking) enables us to directly print microfluidic channels onto the long, thin, implantable needle-like shanks of neural probes while also printing customized fluidic connectors onto the larger base/handle regions of the probe chips. The overall process is summarized in \textbf{Figure \ref{fig:1}A}: starting with 200-mm diameter wafers of foundry-fabricated photonic neural probes, individual probe chips are separated and screened, microfluidic channels and connectors are 3D printed onto the probe chips, and finally, the probe chips are packaged with fluidic tubes and multicore optical fibers. 3D printing microfluidic structures onto existing neural probes is a direct and straightforward path toward extending state of the art photonic and electrophysiological neural probes to include the functionality of neurochemical injection. Importantly, this method avoids modification of the Si microfabrication processes for the neural probes, which, in cases of advanced foundry fabrication process flows, may be prohibitively costly and time-consuming to modify for monolithic integration of microfluidics. Recently, we have developed and demonstrated a variety of foundry-fabricated photonic neural probes with silicon nitride (SiN) waveguides and grating emitters generating addressable low-divergence beams (\cite{sacher2019visible}), light sheets \citep{sacher2021implantable}, and continuously steerable beams \citep{Chen_CLEO_2021, sacher2022optical} for optogenetic stimulation and functional imaging. Additionally, we have integrated microelectrodes onto our photonic neural probes (\textbf{Figure \ref{fig:1}A}, \cite{chen2022implantable}). In this work, we 3D print microfluidic channels onto our photonic neural probes and demonstrate simultaneous photonic and fluidic capabilities.

As an example application utilizing the microfluidic and photonic functionalities of the neural probes, we demonstrate photolysis of a caged compound (uncaging, \textbf{Figure \ref{fig:1}B}). Photoactivatable caged fluorophores or bioactive agents are rendered nonfluorescent or inert by the cage compound, and photolysis by UV or near-UV irradiation cleaves the chemical bond between the cage group and the caged compound, releasing the fluorophores or bioactive agents to become fluorescent or active again. Photoactivatable fluorophores have been widely used in high-resolution photoactivated localization microscopy \citep{hess2009ultrahigh,hauke2017specific}, fluid dynamics studies \citep{fort2021rhodamine}, and molecular transport tracing in biological tissues \citep{martens2004caged,schuster2017photoactivatable}. Photoactivatable neurotransmitters (e.g., caged glutamate) are also widely applied in neuroscience experiments to mimic the timescale and localization of physiological responses of intracellular circuits \citep{trigo2009laser}. Lasers, lamps, and LEDs are most often used as the light sources for uncaging, where photolysis of caged compounds happens at a timescale of milliseconds with milliwatt-scale photoactivation power levels. The dynamic patterned illumination available from our photonic neural probes offers the possibility of deep-brain uncaging with a combination of spatial resolution and low tissue displacement unattainable with conventional tools such as implanted optical fibers and GRIN lenses. In our neural probe uncaging demonstrations, the 3D-printed microfluidic channel injects caged fluorescein and addressable 405-nm wavelength beams emitted by the neural probe selectively uncage small sections of the fluorescein bolus. Uncaging with our neural probes was tested in both solution and fixed brain tissue. Uncaging with photonic neural probes was proposed in \cite{Fowler_2019}, but to our knowledge, our work is the first demonstration of uncaging with a photonic neural probe with multiple addressable emitters and also the first demonstration in brain tissue.

The novel 3D-printed microfluidics integration method and proof of concept demonstrations presented herein open an avenue toward adding fluidic functionalities to the most advanced photonic and electrophysiology Si neural probes available today. This technology offers the potential for new optogenetics experiments leveraging the high-spatiotemporal-resolution illumination and extracellular recording capabilities of state of the art neural probes in addition to the drug injection and neurochemical sampling capabilities of integrated microfluidics. In addition, such neural probes are promising tools for deep-brain uncaging of a variety of photoactivatable compounds, including caged neurotransmitters and fluorophores, achieving a unique combination of spatial resolution and low tissue displacement.

\section{Neural Probe Design, Fabrication, and Packaging}

\subsection{Neural Probe Design}

\begin{figure}[ht]
\begin{center}
\includegraphics[width=1\textwidth]{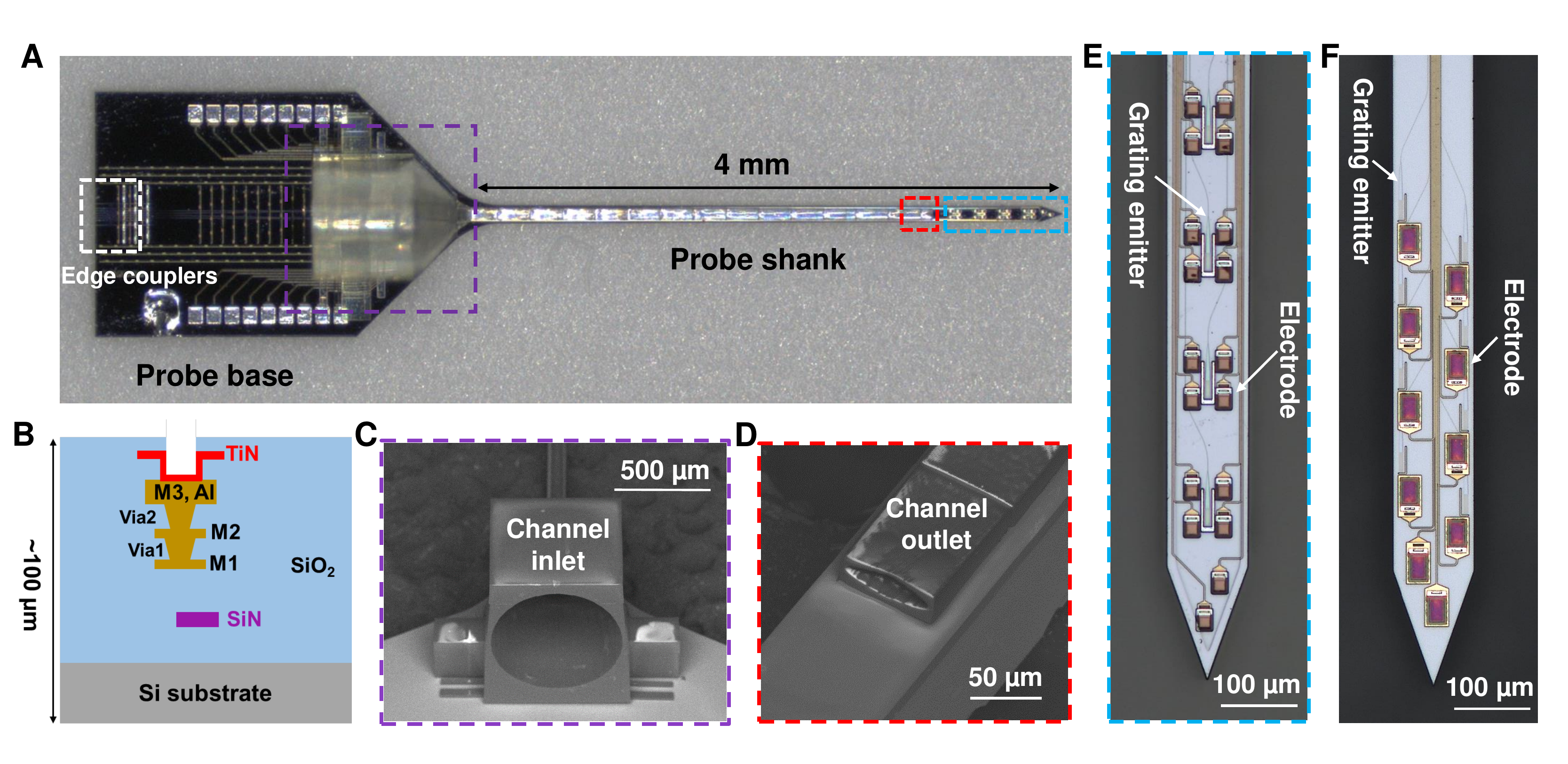}%
\end{center}
\caption{Photonic neural probe with a 3D-printed microfluidic channel. \textbf{(A)} Optical micrograph of the neural probe (Probe 1). \textbf{(B)} Cross-section schematic of the neural probe (microfluidics not shown) with a SiN waveguide layer, 3 Al routing layers, and TiN microelectrodes. SEM images of the \textbf{(C)} inlet and \textbf{(D)} outlet of the 3D-printed microfluidic channel. Optical micrographs of the grating emitters and electrodes on the probe shank of \textbf{(E)} Probe 1 and \textbf{(F)} Probe 2.}
\label{fig:2}
\end{figure}

One of our fabricated photonic neural probes with a 3D-printed microfluidic channel is shown in \textbf{Figure \ref{fig:2}A}. The neural probe consists of a probe base (1.7-mm width and 2.7-mm length) and a 4-mm long implantable shank. The cross-section schematic of the probe chip is shown in \textbf{Figure \ref{fig:2}B} and includes a SiN waveguide layer, 3 aluminum (Al) metal routing layers, and titanium nitride (TiN) microelectrodes; the probe chip thickness is $\approx$100 {\textmu}m. An array of fiber-to-chip edge couplers along the facet of the probe base couples laser light onto the chip from a multicore fiber, routing waveguides direct light up to the shank, and each waveguide is terminated at the distal end of the shank with a grating coupler emitter for light emission (\textbf{Figures \ref{fig:2}E,F}), referred to hereafter as a grating emitter. Each grating emitter is 6 {\textmu}m wide and 25 {\textmu}m long with a 400-nm grating pitch and 50\% duty cycle, resulting in a low-divergence emitted beam, as in \cite{sacher2019visible}. The microelectrodes enable electrophysiological recordings, and we have demonstrated this functionality in \cite{chen2022implantable}. However, in this work, we focus on the photonic and fluidic functionalities, and the microelectrodes are not used. 

The 3D-printed microfluidic channel on the probe has a total length of 4.3 mm, with a channel inlet located on the probe base and a channel outlet on the shank. The channel outer dimensions on the shank are 80 {\textmu}m in width and 30 {\textmu}m in height; the inner dimensions are 70 {\textmu}m in width and 18 {\textmu}m in height. A 1-mm long fluidic connector on the probe base facilitates coupling to an input tube, and the inlet diameter is 750 {\textmu}m. The computer-aided design (CAD) file is included in the Supplementary material. \textbf{Figures \ref{fig:2}C, D} show SEM images of the inlet and outlet of the printed microfluidic channel.

Two probe designs were investigated in this work, referred to hereafter as Probe 1 and Probe 2. Probe 1 (\textbf{Figure \ref{fig:2}E}) exhibited a higher optical transmission at a 473-nm wavelength and was used for characterization of the 3D-printed microfluidic channel (Section \ref{channel characterization}) and optical beam profiles (Sections \ref{optical characterization}, \ref{simultaneous}). Probe 2 (\textbf{Figure \ref{fig:2}F}) had a higher optical transmission at a 405-nm wavelength and was used for the uncaging experiments (Sections \ref{uncage in fluorescein}, \ref{uncage in brain}). Additional differences between Probes 1 and 2 include (Supplementary \textbf{Figure S1}): 1) fabrication on different wafers, 2) Probes 1 and 2 had nominal SiN thicknesses of 200 nm and 150 nm, respectively, 3) Probe 2 had an extra SiN waveguide layer for defining bi-layer edge couplers, as in \cite{lin2021low} and \cite{Sacher_2023}, and 4) Probe 1 had a single column of four grating emitters (170 {\textmu}m pitch) and Probe 2 had 2 columns of grating emitters (16 emitters in total, 85 {\textmu}m and 110 {\textmu}m lateral and longitudinal inter-grating pitches, respectively).

\subsection{Neural Probe Fabrication}
The implantable neural probes were fabricated on 200-mm diameter silicon wafers at Advanced Micro Foundry (\textbf{Figure \ref{fig:1}A}). First, the silicon dioxide (SiO$_2$) bottom cladding of the waveguides was deposited by plasma enhanced chemical vapor deposition (PECVD). Next, SiN waveguides were defined by PECVD, 193-nm deep ultraviolet photolithography, and reactive ion etching. The SiO$_2$ top cladding, Al routing layers, and TiN microelectrodes were formed next. Chemical mechanical polishing was used for layer planarization. Finally, deep trenches were etched into the wafers to define the probe shapes and edge coupler facets, and the wafers were thinned to $\approx$100 {\textmu}m by backgrinding.

\subsection{Two Photon Polymerization 3D Printing of Microfluidic Channels}
\begin{figure}[ht]
\begin{center}
\includegraphics[width=1\textwidth]{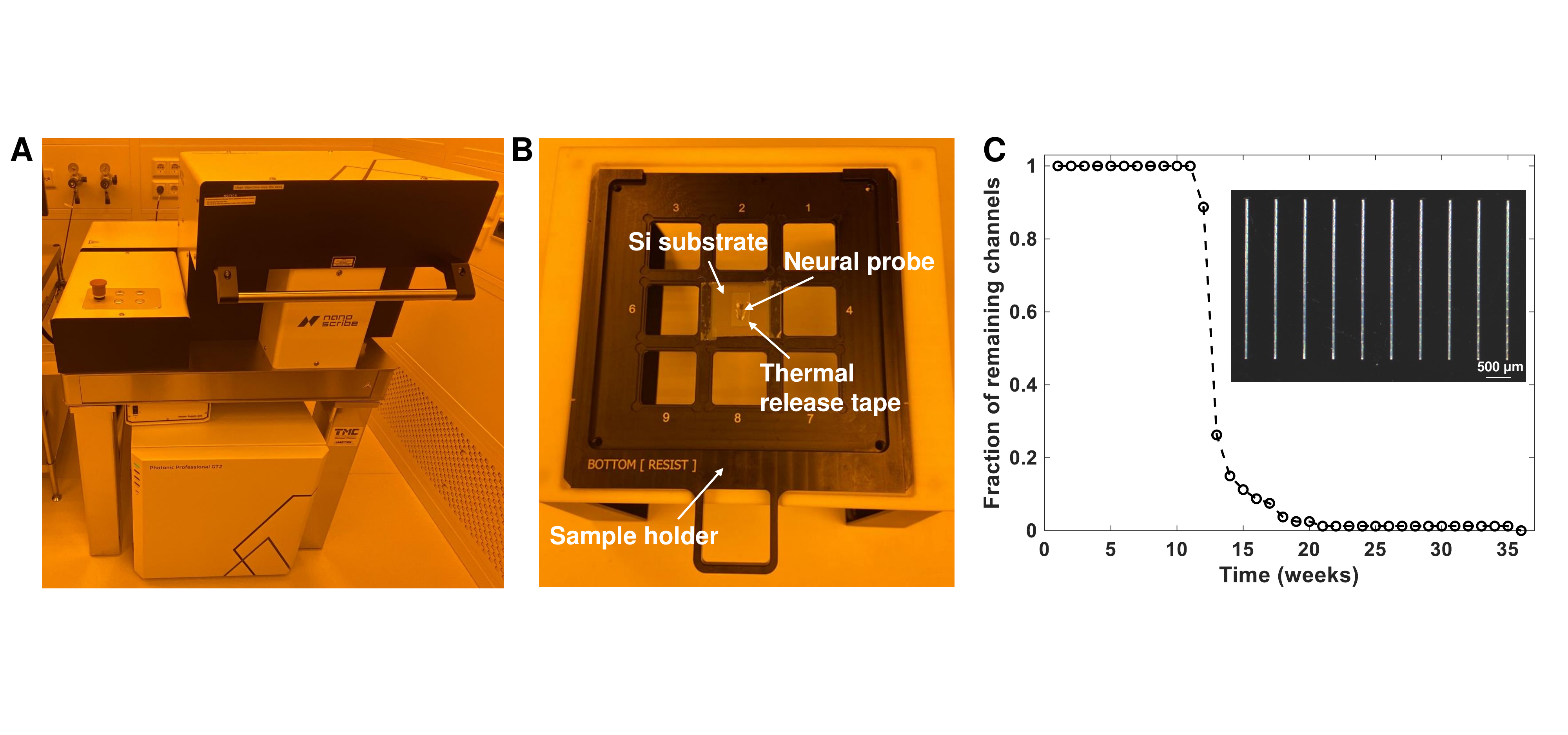}%
\end{center}
\caption{Two photon polymerization 3D printing of the microfluidic channels. \textbf{(A)} Photograph of the 3D printing tool (Nanoscribe Photonic Professional GT2 laser writer). \textbf{(B)} Photograph of a neural probe chip mounted on the sample holder of the 3D printing tool. \textbf{(C)} Fraction of printed channels (4x20 array, 3-mm long) remaining on a Si substrate after immersion in artificial cerbrospinal fluid (aCSF) at \SI{37}{\celsius} for multiple weeks. The inset shows an optical micrograph of a portion of the channel array on the substrate.}
\label{fig:3}
\end{figure}

Following fabrication of the neural probe wafers, individual neural probe chips were screened and selected for 3D printing of microfluidic channels at the Max Planck Institute of Microstructure Physics. A two-photon laser writer (Photonic Professional GT2, Nanoscribe GmbH, Karlsruhe, Germany) was utilized for highly flexible and precise additive printing of microfluidic structures onto neural probes (\textbf{Figure \ref{fig:3}A}). The Nanoscribe 3D writer enables layer-by-layer structural formation based on two-photon polymerization. A 780 nm femtosecond laser is used as the laser source during writing, and various feature sizes can be achieved with the combination of different objective lenses and resins. The microfluidic structures were directly written on the neural probes. Before printing, neural probes were pretreated by silanization to enhance the surface adhesion of the printed structures. After rinsing with acetone, isopropanol, and deionized water in sequence, clean and dry neural probes were soaked in a silane solution overnight. The silane solution was prepared by mixing acetone with (3-glycidyloxypropyl)trimethoxysilane (GPTMS, Sigma-Aldrich, St. Louis, MO, USA) at a ratio of 99:1. Next, the neural probes were mounted on Si chips as shown in \textbf{Figure \ref{fig:3}B}; a silicon wafer was cut into 25 mm x 25 mm chips to match the grid of the sample holder of the two-photon laser writer, double-sided thermal release tape (RA-95L(N), TELTEC GmbH, Mainhardt, Germany) was cut to a suitable size and applied onto the silicon chips, and after taking the neural probes out from the silane and rinsing with deionized water, the dried probes were loaded onto the front side of the thermal release tape (one probe per Si chip). We used a 25x immersion objective (LCI Plan-Neofluar 25x/0.8 Imm Corr DIC M27, Carl Zeiss Microscopy Deutschland GmbH, Oberkochen, Germany) in combination with the photoresin IP-S (Nanoscribe GmbH) for printing of the designed microfluidic structures. The dip-in laser lithography configuration was used during printing. For microfluidic structures with the aforementioned geometry, the printing time was approximately 2.5 hours per probe.

Following completion of the laser writing procedure, the sample holder was taken out of the printing chamber. To release the neural probes with printed structures, each silicon chip was placed onto a hotplate (VWR International, LLC., Dresden, Germany) preheated to \SI{100}{\celsius}. Each neural probe was released from the thermal tape after heating for one minute. Then the probes were soaked in a propylene glycol monomethyl ether acetate (PGMEA, 484431, Sigma-Aldrich) developer bath for development of the laser-written structures. Since the IP-S resin is a negative tone photoresist, the excess resin must be removed to reveal the final microfluidic structure. Here, development at \SI{40}{\celsius} with magnetic stirring at 150 rpm for 3 hours fully washed away the unexposed resin inside the microfluidic channels. After development and cleaning, the microfluidic integration process was complete and the probes were packaged and characterized. To validate the adhesion strength of the 3D-printed structures in experimental conditions, a 4x20 array of 3-mm long microfluidic channels was fabricated on a 25 mm x 25 mm silicon substrate following the above workflow. The silicon substrate was placed into a vial filled with artificial cerebrospinal fluid (aCSF; 92 mM NaCl, 25 mM glucose, 20 mM HEPES, 2.5 mM KCl, 2 mM MgSO$_4$, 2 mM CaCl$_2$, 1.2 mM NaH$_2$PO$_4$; pH 7.3-7.5), and the vial was sealed and kept inside a cell culture incubator (Binder GmbH, Tuttlingen, Germany) with 5\% CO$_2$ and humidified atmosphere at \SI{37}{\celsius}. \textbf{Figure \ref{fig:3}C} shows the fraction of channels remaining on the substrate after multiple weeks. All printed channels stayed adhered to the silicon substrate after immersion in aCSF for 10 weeks. The detachment of printed channels happened from the 11th week, and the last channel delaminated from the substrate during the 36th week.

\subsection{Neural Probe Packaging and Grating Emitter Addressing}
The photonic neural probes with integrated microfluidic channels were packaged onto holders with multicore optical fibers and fluidic tubes for deployment in experiments. \textbf{Figure \ref{fig:4}A} shows a photograph of Probe 1 after packaging. The packaging procedure for each neural probe is outlined below.

First, the neural probe was glued onto a 3D-printed probe holder with 5-min epoxy (Loctite, 1365868, Westlake, OH, USA). Controlled by a translation stage (MAX311D, Thorlabs, Newton, NJ, USA), a \ang{90}-bent rigid metal adaptor tube attached to polytetrafluoroethylene (PTFE) tube (0.5 mm inner diameter, 1 mm outer diameter, S1810-08, Bohlender GmbH, Germany) was inserted into the microfluidic channel inlet. The metal adaptor had a 330 {\textmu}m inner diameter and 640 {\textmu}m outer diameter to match the inlet size of the printed microfluidic structure. A stable and leak-free fluidic connection was achieved by surrounding the connection site with UV-curable epoxy (Katiobond GE680, Delo, Germany) and completely curing with a UV curing light-emitting diode (LED) system (CS2010, Thorlabs). Following attachment of the fluidic tube, optical fiber attachment was performed. To enable addressing of the multiple grating emitters on the neural probe, a custom visible-light 16-core optical fiber \citep{azadeh2022multicore} (fixed in a fiber ferrule) was actively aligned to the array of input edge couplers on the probe chip and permanently attached using UV-curable low-shrinkage epoxies (OP-67-LS and OP-4-20632, Dymax Co., Torrington, CT, USA). The alignment was performed with a 5-axis piezo fiber alignment stage and top-down microscope. The optical emission power from the grating emitters was monitored during the epoxy application process. Following UV-curing of the low-shrinkange epoxies, the optically-opaque epoxy (EPO-TEK 320, Epoxy Technology, Billerica, MA, USA) was spread onto the outer surface of the epoxy encapsulation to block any stray light that was not coupled to the on-chip waveguides. A steel mounting rod was attached to the probe holder via a steel screw. Finally, the packaged neural probe was mounted on a 4-axis micromanipulator (uMp-4, Sensapex, Oulu, Finland) via the steel rod.

To address the grating emitters on the packaged neural probe, the distal end of the multicore fiber (MCF) was connected and aligned to a custom, computer-controlled, free-space optical scanning system based on a microelectromechanical system (MEMS) mirror \citep{Wahn_2021, chen2022implantable}. An input laser beam coupled to the scanning system was directed and coupled to a core of the MCF (selected by the MEMS mirror position), guided to the probe chip via the MCF, coupled onto the chip via the corresponding edge coupler on the probe, guided down the shank by a routing waveguide, and emitted by the corresponding grating emitter. The various MCF cores and hence grating emitters were addressed by controlling the MEMS mirror position. In principle, the time required for switching between grating emitters can be as small as the MEMS mirror response time ($\approx$ 5 ms), but in practice, communication between the computer and MEMS mirror controller adds latency.   
\label{probe packaging and addressing}

\section{Materials and methods}

\subsection{Preparation of Agarose Gel Block}
To prepare the agarose gel blocks, 100 mg of agarose powder (Ultrapure\texttrademark, Invitrogen, Waltham, MA, USA) was mixed with 10 mL of Milli-Q water to form a 1\% w/v agarose solution. The solution was heated until boiled and then poured into a plastic mold and stored in a refrigerator until solidified. 

\subsection{Microfluidic Channel Characterization}
The performance of the 3D-printed microfluidic channel on Probe 1 was characterized (Section \ref{channel characterization}). A 4 mM Allura red dye solution (A0943, TCI Deutschland GmbH, Eschborn, Germany) was prepared to investigate the diffusion profile at the channel outlet \citep{wang2022colorimetry}. The packaged neural probe was fixed to a translation stage for insertion into a 1\% w/v agarose gel block. Two USB cameras were used to capture the diffusion profile simultaneously from both the front and side of the neural probe shank; the probe, agarose gel block, and cameras were aligned such that the side and front views passed through flat interfaces of the agarose gel. Precise flow control at a fixed flow rate was achieved with a syringe pump (Pump 33 DDS, Harvard Apparatus, Holliston, MA, USA), and red dye was injected into the agarose gel block through the 3D-printed microfluidic channel at a flow rate of 10 {\textmu}L/min. Flow resistance of the channel was also characterized with the channel outlet exposed to air and injection of deionized water. The injection was driven by the syringe pump, and a pressure sensor (MPS4, Elveflow, Elvesys, Paris, France) was connected inline to record the pressure in real-time.
\label{flow resistance}

\subsection{Beam Profile Characterization and Fluorescence Imaging Apparatus}
Probe 1 (\textbf{Figure \ref{fig:2}E}) was used for beam profile characterization in a fluorescein solution (Section \ref{optical characterization}). Light from a 473-nm laser (OBIS, LX 473-50, Coherent Inc., Santa Clara, CA, USA) was coupled onto the neural probe via the optical scanning system and MCF, as described in Section \ref{probe packaging and addressing}, and light was emitted from the grating emitters into the solution. As shown in \textbf{Figure \ref{fig:4}E}, the packaged neural probe was mounted on a micromanipulator and immersed in a 100 {\textmu}M fluorescein sodium salt solution bath (46960-100G-F, Sigma-Aldrich). The probe was angled such that the emitted beams were parallel to the surface of the solution. The input polarization state to the scanning system was adjusted using an inline fiber polarization controller to achieve either a transverse-electric or transverse-magnetic polarization in the on-chip waveguides (to avoid emission of 2 beams from each grating emitter). The scanning system was controlled via a computer interface to select the grating emitter for each beam profile measurement.

Beam profiles were collected via fluorescence imaging. The 473-nm emission of the neural probe excited the fluorescein, and microscopes aligned above and at the side of the bath imaged the resultant fluorescence. The side-view beam profiles were imaged through a flat transparent window in the fluorescein chamber, and the side-view microscope was equipped with a 5x objective (M Plan Apo 5x, NA=0.14, Mitutoyo Deutschland GmbH, Berlin, Germany) and a CMOS camera (Grasshopper3, USB 3.0, Teledyne FLIR, Wilsonville, OR, USA) with an emission filter (MF525-39, Thorlabs) installed in front to reject excitation light from the probe. The top-down beam profiles were imaged by a wide-field epifluorescence microscope (Cerna, Thorlabs) with a 10x objective (M Plan Apo 10x, NA=0.28, Mitutoyo Deutschland GmbH), an EGFP filter cube (49002, Chroma Technology Corporation, Bellows Falls, VT, USA), and an sCMOS camera (Prime BSI, Teledyne, Photometrics, Tucson, AZ, USA). The top-down epifluorescence microscope was also used for imaging of uncaged fluorescein in Sections \ref{uncage in fluorescein} and \ref{uncage in brain}, and an LED (pE-4000, CoolLED, Andover, UK) coupled to the microscope was used for epi-illumination.

\label{beam profile}

\subsection{Preparation of Caged Fluorescein Solution}
Fluorescein bis-(5-carboxymethoxy-2-nitrobenzyl) ether, dipotassium salt (CMNB-caged fluorescein, Invitrogen F7103, Invitrogen) was used for the validation of photolysis with the neural probe at 405 nm (Sections \ref{uncage in fluorescein}, \ref{uncage in brain}). A 100 {\textmu}M CMNB-caged fluorescein solution was prepared by dissolving the dye in 1x phosphate-buffered saline (PBS, diluted from 10x concentrated PBS, Sigma-Aldrich) and stored in a refrigerator at \SI{4}{\celsius} in darkness. The caged dye solution was filtered with syringe filters (5 {\textmu}m pore size, Mimisart, Sartorius, Germany) before experiments.

\subsection{Preparation of Fixed Brain Tissue}
Fixed brain tissues were prepared from wild-type mice (female, age $\approx$60-90 days). The mice were euthanized with CO$_2$ and transcardially perfused with 1x PBS followed with 1.5\% paraformaldehyde (PFA). After extraction, the whole brain was kept in 1.5\% PFA at \SI{4}{\celsius} for 12 h for fixation. The whole brain was then kept in 1x PBS at \SI{4}{\celsius} after fixation. Before experiments, 2-mm thick coronal brain slices were prepared with a brain matrix (Alto Brain Matrix stainless steel 1 mm mouse coronal 45-75 gm; Harvard Apparatus) and stirrup-shaped blades (Type 102, Carl Roth GmbH + Co. KG, Karslruhe, Germany).
\label{brain slice preparation}

\section{Results}

\subsection{Neural Probe Characterization}\label{probe characterization}

The optical and fluidic functionalities of Probe 1 were characterized in detail. Probe 2 was directly used in uncaging experiments and the full characterization procedure was not performed. However, tests to screen Probe 2 prior to uncaging experiments indicated similar beam profiles and microfluidic channel properties to Probe 1. In this section, Probe 1 characterization results are summarized.  

\subsubsection{Microfluidic Channel Characterization}\label{channel characterization}
\begin{figure}[ht]
\begin{center}
\includegraphics[width=1\textwidth]{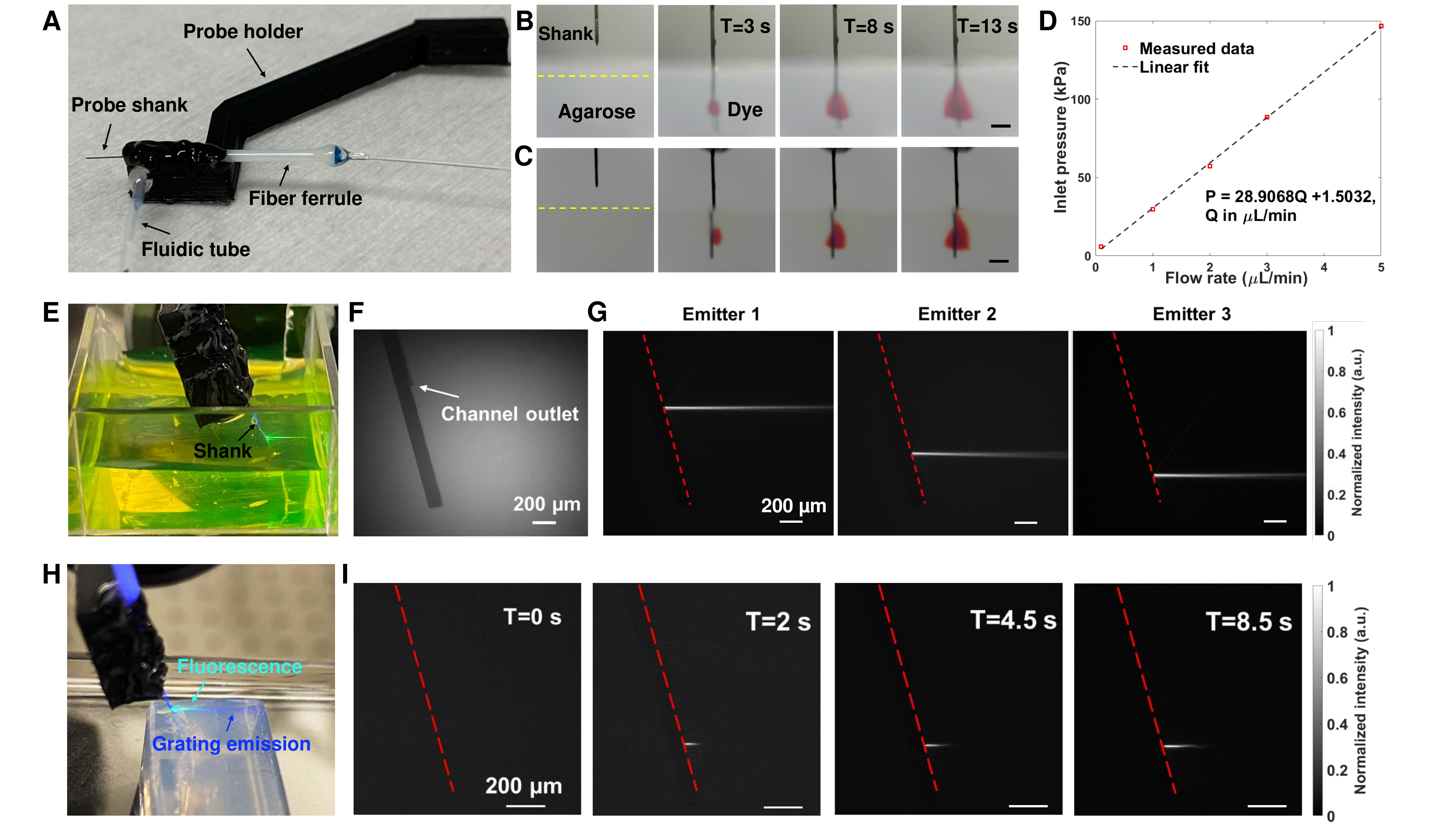}%
\end{center}
\caption{Fluidic and optical characterization of Probe 1. \textbf{(A)} Photograph of the packaged neural probe. \textbf{(B)} Front-view and \textbf{(C)} side-view sequential photographs of red dye diffusion in 1\% w/v agarose gel with injection from the probe at a flow rate of 10 {\textmu}L/min. The scale bars are 500 {\textmu}m. \textbf{(D)} Measured inlet pressure P vs. flow rate Q. \textbf{(E)} Photograph of the packaged probe immersed in a fluorescein solution with illumination from one of the grating emitters (raw image mirrored for clarity). \textbf{(F,G)} Grey-scale side-view micrographs of the probe shank immersed in a fluorescein solution. \textbf{(F)} Grating emitters off and ambient illumination applied (for visibility of the shank). \textbf{(G)} Fluorescence imaging of the side beam profiles of various grating emitters (grating emitters on, ambient illumination off). The dashed line indicates the top surface of the probe shank. \textbf{(H)} Photograph of the probe inserted into a block of 1\% w/v agarose gel while emitting light and injecting fluorescein (raw image mirrored for clarity). \textbf{(I)} Side-view fluorescence images of the beam profile evolution in agarose gel as the injected fluorescein diffuses (grating emitter on); the beam length increases with time as the fluorescein diffuses and the overlap between the beam and fluorescein increases.}
\label{fig:4}
\end{figure}

Following the procedure in Section \ref{flow resistance}, the neural probe was inserted into a block of agarose gel, red dye was injected by the microfluidic channel at 10 {\textmu}L/min, and the diffusion profile of the red dye was imaged by cameras aligned to the front and side of the probe shank. Sequential photographs of the red dye diffusion inside the agarose gel after 3 s, 8 s, and 13 s are shown in \textbf{Figures \ref{fig:4}B} and \textbf{\ref{fig:4}C}. After 3 s, the red dye diffusion profile had transverse and axial extents of $\sim$0.33 mm and $\sim$0.36 mm, respectively, covering 2 grating emitters on Probe 1. After 13 s, the axial extent increased to about 0.6 mm, covering 3 of the 4 grating emitters on the shank. A front-view video of the dye diffusion is shown in Supplementary Video 1. The neural probe was inserted around 1.5 mm deep into the agarose gel such that the dye diffusion profile was not cropped by the top surface of the agarose gel block during the period of observation. After insertion and retraction from the agarose gel, the 3D-printed microfluidic channel remained attached to the probe shank with no signs of delamination, indicating that the microfluidic channel had sufficient adhesion to the probe shank and mechanical strength against shear force for the insertion into agarose gel. Following Section \ref{flow resistance}, channel inlet pressure vs. flow rate was measured with the probe in air and injection of deionized water (\textbf{Figure \ref{fig:4}D}), and a linear relationship was observed, similar to \cite{lee2015multichannel} and \cite{shin2019multifunctional}.

\subsubsection{Optical Characterization}\label{optical characterization}
The optical transmission and beam profiles of Probe 1 were measured. We define the transmission to be that of the optical scanning system, MCF, and packaged neural probe together as a system, i.e., the ratio of emitted optical power from the grating emitters and the input laser power to the optical scanning system. The transmission at a wavelength of 473 nm varied from $\num{-27}$ to $\num{-20}$ dB (median of about $\num{-22}$ dB) across the grating emitters on the probe, and the large variation in transmission was attributed to alignment drift during attachment of the MCF to the probe chip facet. The transmission variation can be compensated by modulating the input laser power or adjusting the MEMS mirror positions for each grating emitter for power equalization.

As described in Section \ref{beam profile}, top and side beam profiles of the neural probe were imaged in a fluorescein solution; the 473-nm emission from the neural probe excited the fluorescein and fluorescence microscopes above and at the side of the probe imaged the resultant fluorescence. \textbf{Figure \ref{fig:4}F} shows the side view of the probe shank in the fluorescein solution without optical emission, and a segment of the microfluidic channel is visible on the probe shank. The side beam profiles in fluorescein from 3 of the grating emitters are shown in \textbf{Figure \ref{fig:4}G}, and the average emission angle is about \ang{19} (relative to the normal of the shank). Top-down beam profiles were also measured, and the average full width at half maximum (FWHM) of the beam after a 300 {\textmu}m propagation distance was about 37 {\textmu}m (Supplementary \textbf{Figure S2}). Beam profiles in agarose gel were also characterized by adding fluorescein into the agarose gel block. The average emission angle was measured to be \ang{16} in agarose gel and the average FWHM after propagating for 300 {\textmu}m was 33 {\textmu}m (Supplementary \textbf{Figure S2}), similar to the beam profile data collected in the fluorescein bath. The low-divergence, neuron-scale beams from the grating emitters are a critical feature of these photonic neural probes, enabling targeted optogenetic stimulation and, as shown in Sections \ref{uncage in fluorescein} and \ref{uncage in brain}, targeted photolysis of caged compounds.

\subsubsection{Simultaneous Fluid Injection and Light Emission}
As a demonstration of simultaneous light and fluid delivery, the neural probe was inserted into an agarose gel block and a 100 {\textmu}M fluorescein solution was injected through the microfluidic channel at a flow rate of 10 {\textmu}L/min. One of the grating emitters was addressed by the scanning system (``switched on") before the injection of the fluorescein solution. \textbf{Figure \ref{fig:4}H} shows a photograph of the probe inserted into the agarose gel with a grating emitter on following infusion and diffusion of the fluorescein solution; fluorescence resulting from the excitation of the injected fluorescein by the grating emitter illumination is apparent in the cyan region of the photograph (a mixture of the blue excitation and green emission light). Top-down and side-view beam profiles in the agarose gel were imaged using microscopes aligned to the top and side of the agarose gel block; the same microscopes used for beam profile measurements in the fluorescein bath (Section \ref{beam profile}). \textbf{Figure \ref{fig:4}I} shows the beam profile evolution at different times following the start of the fluorescein solution injection. As the infused fluorescein diffused in the agarose gel, a larger volume of fluorescein was excited by the grating emission. It is noteworthy that the beam intensity remained stable during the infusion of the fluorescein solution, indicating a steady delivery of both light and fluid. A top-down view video of simultaneous light emission and fluid delivery is shown in Supplementary Video 2.
\label{simultaneous}

\subsection{Uncaging in a Caged Fluorescein Solution}
As an example application of the photonic and fluidic functionalities of the neural probes, Probe 2 (\textbf{Figure \ref{fig:2}F}) was used in a series of uncaging experiments. As a proof of concept, the uncaging experiments were designed to demonstrate the photolysis of a caged fluorophore with optical emission from the neural probe. Uncaging of a fluorophore enables a simple quantification of the uncaging via fluorescence imaging. The luminescence of a photoactivatable fluorophore is prevented by cage groups, which can be removed from the fluorophore molecule by photocleavage of the chemical bonds. The peak absorption wavelength of cage groups is typically UV or near-UV \citep{fort2021rhodamine}. Here, we focus on CMNB-caged fluorescein, which has two CMNB cage groups bonded to the fluorescein molecule, and photocleavage of the cages at 405 nm allows fluorescence emission upon excitation with blue light \citep{chan2016droplay}.

\textbf{Figure \ref{fig:5}A} shows the imaging apparatus used in the uncaging experiments. The packaged neural probe was mounted on a micromanipulator for precise control over the orientation and position of the neural probe. A wide-field epifluorescence microscope with a long working distance 10x objective (described in Section \ref{beam profile}) was positioned above the probe and sample for fluorescence imaging of the uncaged fluorescein. A 405-nm fiber-coupled laser (OBIS, LX 405-100, Coherent Inc., Santa Clara, California) was coupled to the optical scanning system, and the resultant 405-nm beams from the grating emitters on the neural probe were used for photoactivation.

Throughout the various uncaging experiments, the imaging apparatus remained unchanged with the exception of the probe orientation and the sample. In addition, the photoactivation and fluorescence imaging protocol remained largely the same, with small modifications to the photoactivation pulse widths.

\label{uncage in fluorescein}

\subsubsection{Sequential Photolysis of Caged Fluorescein in Solution}

\begin{figure}[ht]
\begin{center}
\includegraphics[width=1\textwidth]{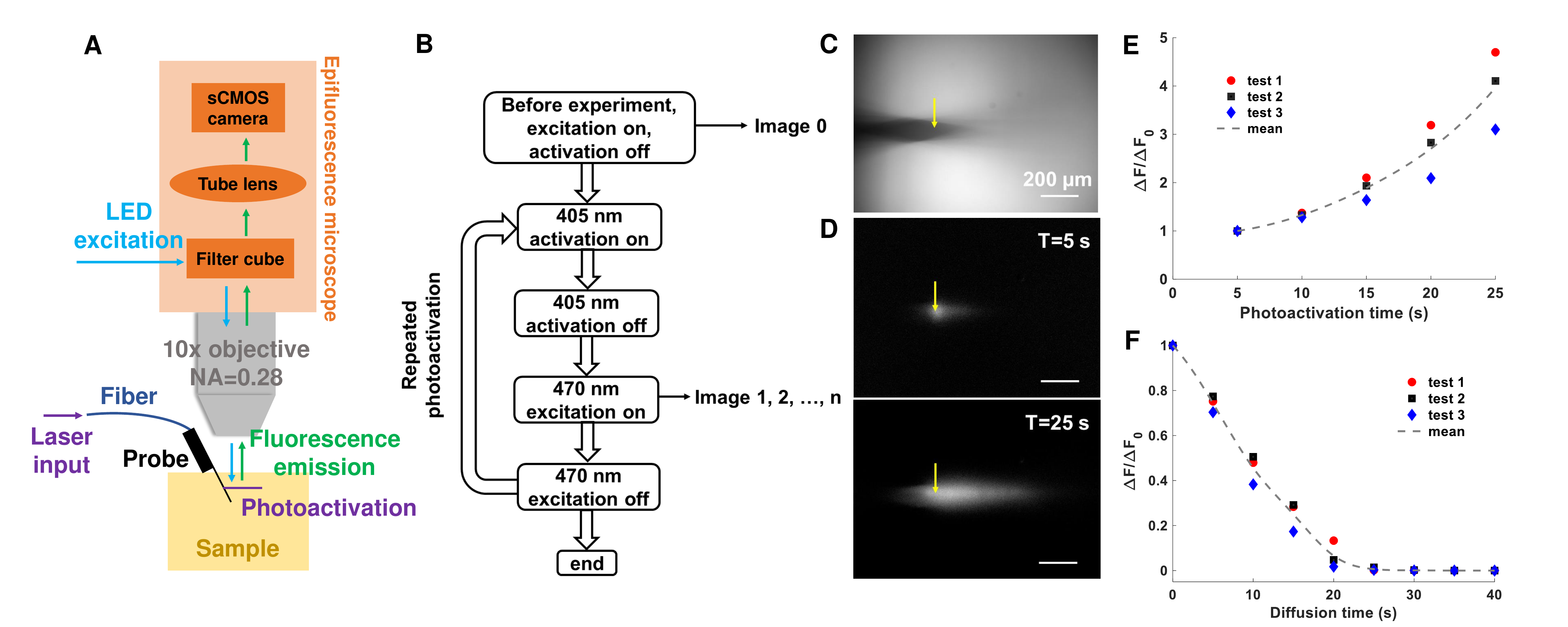}  
\end{center}
\caption{Uncaging with a neural probe (Probe 2) in a CMNB-caged fluorescein solution bath. \textbf{(A)} Illustration of the experimental setup (not to scale); light emitted from the neural probe (405-nm wavelength) uncaged the fluorescein, and a widefield fluorescence microscope above the bath imaged the uncaged fluorescein profiles (470-nm wavelength LED excitation). \textbf{(B)} Flow chart of the uncaging experiment workflow. First, a reference image was collected, and next, a series of repeated photoactivations were applied. For each photoactivation, a 5 s optical pulse was emitted by the probe (15 {\textmu}W grating emission power), and subsequently, a fluorescence image was collected using the fluorescence microscope. \textbf{(C)} Top-view fluorescence image of the probe shank in the caged fluorescein solution prior to photoactivation (reference image). The grating emitter used in this experiment is delineated by the yellow arrow. \textbf{(D)} Top-view fluorescence profiles with background subtraction after \textbf{(top)} 1 and \textbf{(bottom)} 5 photoactivation pulses. The scale bars are 200 {\textmu}m. \textbf{(E)} Normalized progressive fluorescence intensity change \textit{$\Delta$F}/\textit{$\Delta$F$_0$} with sequential photoactivation. \textit{$\Delta$F} is the mean pixel intensity in the selected ROI after background subtraction, and \textit{$\Delta$F$_0$} is the mean pixel intensity after the first photoactivation pulse. \textbf{(F)} Normalized fluorescence intensity change \textit{$\Delta$F/\textit{$\Delta$F$_0$}} during diffusion of the uncaged fluorescein after the photoactivation pulses in \textbf{(E)}, i.e., the last data points in \textbf{(E)} correspond to the first of \textbf{(F)} and no photoactivation light is applied in \textbf{(F)}. The dotted curves show the average of three experiments.}
\label{fig:5}
\end{figure}

In the first uncaging experiment, the photonic functionality of the neural probe was tested separately from the microfluidics. The probe shank was immersed in a chamber with 3 mL of 100 {\textmu}M CMNB-caged fluorescein solution. The probe was mounted at an angle of \ang{27} to ensure the 405-nm emitted beam was parallel to the focal plane of the microscope, avoiding some portions of the beam (and resulting fluorescence) being out of focus during fluorescence imaging. \textbf{Figure \ref{fig:5}B} is a block diagram describing the sequential workflow of the uncaging experiment. At the beginning of the experiment, an image with 470-nm excitation from the microscope (epi-illumination, intensity $\sim$174 mW/cm\textsuperscript{2}) was captured as a reference for subsequent measurements of the fluorescence intensity change (\textit{$\Delta$F}) due to photolysis. One of the grating emitters was switched on for a predetermined photoactivation period (5-s duration, 15 {\textmu}W grating emission power), and the epi-illumination was turned off during this step. After completion of the photoactivation step, the epi-illumination was immediately turned on and a fluorescence image was acquired; each image provides a measure of \textit{$\Delta$F} due to photoactivation (following subtraction of the reference image). The epi-illumination was turned off immediately after the image acquisition, and the grating emitter was switched on again for the next photoactivation. The photoactivation and fluorescence imaging steps were repeated 5 times to measure the sequential \textit{$\Delta$F} caused by photolysis at 405 nm (for a cumulative photoactivation time of 25 s). The grating emitter was always switched off during fluorescence imaging with epi-illumination, and the epi-illumination was switched off during during photoactivation with the grating emitter. The alternating photoactivation and epi-illumination, as well as the image acquisition, were automated using Matlab R2021a (Mathworks, Natick, MA, USA). Each gap between photoactivation periods (required for fluorescence image collection) was about 0.98 s. The exposure time for the sCMOS camera attached to the fluorescence microscope was set to 100 ms, and the epi-illumination power (intensity $\sim$174 mW/cm\textsuperscript{2}) was chosen for reasonable signal to noise ratios in the fluorescence images. The optical transmission of Probe 2 (coupled to the optical scanning system) at 405 nm varied from $\num{-31}$ to $\num{-24}$ dB, with a median of about $\num{-27}$ dB, across the grating emitters. The transmission and laser power limited the output power of the neural probe and the minimum photoactivation time durations.

\textbf{Figure \ref{fig:5}C} shows a top-down fluorescence image of the neural probe shank immersed in the CMNB-caged fluorescein solution (epi-illumination on) prior to photoactivation. The image was used as a reference for background subtraction in subsequent fluorescence images after photoactivation. The position of the neural probe was adjusted to ensure the target optical emitter was within the focal plane (delineated by the yellow arrow). \textbf{Figure \ref{fig:5}D} shows the time-dependent distribution of the uncaged fluorescein (fluorescence images with background subtraction) with accumulated photoactivation of 5 s and 25 s in the top and bottom images, respectively. With an increasing fluence of 405-nm grating emission, the number of free-form fluorescein molecules increased and led to stronger fluorescence emission. \textbf{Figure \ref{fig:5}E} shows \textit{$\Delta$F}/\textit{$\Delta$F$_0$} vs. photoactivation time, and \textit{$\Delta$F} is the mean pixel intensity in the selected region of interest (ROI) (calculated using Fiji, \cite{Schindelin_2012}); the ROI is shown in Supplementary \textbf{Figure S3} and was chosen to contain the diffusion profile of the uncaged fluorescein. \textit{$\Delta$F$_0$} is the mean pixel intensity in the ROI after the first photoactivation pulse, and \textit{$\Delta$F}/\textit{$\Delta$F$_0$} progressively increased during the sequential photoactivations. Three tests were repeated with the same parameters, and the solution was stirred between tests to prevent disturbance of the measurements by uncaged fluorescein from the previous test.
\label{sequential}

\subsubsection{Characterization of the Diffusion of Uncaged Fluorescein}\label{diffusion}

To investigate the diffusion of the uncaged fluorescein in the CMNB-caged fluorescein bath, a sequence of images was taken after the 5th photoactivation. The 470-nm epi-illumination from the microscope was briefly turned on every 5 s and a fluorescence image was acquired. The epi-illumination was turned off between image acquisitions, and the process was automated using Matlab R2021a. Identical epi-illumination power and camera exposure time to Section \ref{sequential} were used. \textbf{Figure \ref{fig:5}F} shows the normalized fluorescence intensity change (\textit{$\Delta$F/\textit{$\Delta$F$_0$}}) during the diffusion of the uncaged fluorescein. The experiment was repeated 3 times. The fluorescence intensity dropped rapidly as the uncaged fluorescein diffused in the solution, and the fluorescence intensity was nearly zero within the field of view of the microscope after 25 s of diffusion, with a 1/\textit{e} decay time of $\sim$10.2 s. During the uncaging experiment in the caged fluorescein solution, the fast diffusion of released fluorescein molecules competed with the fluorescence intensity increase due to photoactivation, and thus, a high fluence of 405-nm light was required for the observation of significant fluorescence intensity increases.

\subsubsection{Localized and Addressable Photolysis at 405 nm}
\begin{figure}[ht]
\begin{center}
\includegraphics[width=1\textwidth]{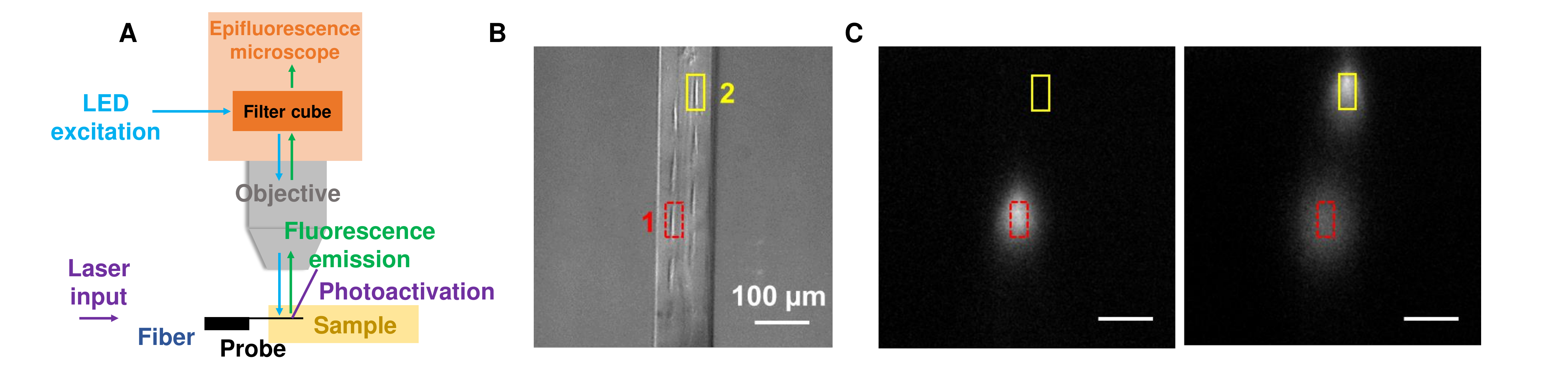}%
\end{center}
\caption{Local, addressable photolysis with a neural probe (Probe 2) in a CMNB-caged fluorescein solution. \textbf{(A)} Illustration of the experimental setup; the probe was inserted into the sample (droplet of solution sandwiched between two coverslips), 405-nm wavelength beams from the addressable grating emitters uncaged the fluorescein, and a widefield fluorescence microscope above the sample (as in \textbf{Figure \ref{fig:5}A}) imaged the uncaged fluorescein. \textbf{(B)} Top-view optical micrograph of the probe shank in the caged fluorescein solution (filter cube removed for visibility of the shank). The locations of grating emitters 1 and 2 are delineated by the red and yellow boxes, respectively. \textbf{(C)} Fluorescence profiles after photoactivation with emitters 1 and 2 (2 s optical pulse, 15 {\textmu}W grating emission power). \textbf{(Left)} Photoactivation from emitter 1 was applied first and a fluorescence image was collected. \textbf{(Right)} Photoactivation from emitter 2 was then immediately applied and another fluorescence image was acquired. The scale bars are 100 {\textmu}m. The distinct locations of the fluorescence spots are indicative of localized, addressable photoactivation from the neural probe.}
\label{fig:6}
\end{figure}

In a second uncaging experiment, we demonstrate the high spatial selectivity of photolysis with grating emitters on the neural probe. The imaging setup was modified as shown in \textbf{Figure \ref{fig:6}A}. The neural probe was mounted horizontally on the micromanipulator so the difference in the locations of grating emitters, and hence photoactivation spots, was maximized in the fluorescence images acquired by the microscope. A droplet of CMNB-caged fluorescein was sandwiched between two coverslips, and the probe shank was inserted into the droplet through the gap between the coverslips, similar to the method in \cite{Segev_2016}. \textbf{Figure \ref{fig:6}B} shows the top-down view brightfield image of the probe shank in the caged fluorescein solution, and eight grating emitters are visible. Two grating emitters were switched on in sequence, and their locations are delineated in \textbf{Figure \ref{fig:6}B}. First, emitter 1 was switched on for 2 s with 15 {\textmu}W of emission power. Next, emitter 1 was switched off and a fluorescence image was acquired. Emitter 2 was then switched on for 2 s with 15 {\textmu}W of emission power, and a second fluorescence image was captured. \textbf{Figure \ref{fig:6}C} shows the fluorescence profiles after photoactivation from grating emitter 1 (left panel) and 2 (right panel). One spot was present after photoactivation from grating emitter 1, two distinguishable spots were present after photoactivation from grating emitter 2, and the location of the spots corresponded to those of the emitters, indicating localized photoactivation. The spot corresponding to grating emitter 1 becomes darker and larger in the second panel of \textbf{Figure \ref{fig:6}C}, a result of the diffusion of the uncaged fluorescein molecules.  

It is noteworthy that the custom 16-core fiber is single-mode over most of the visible spectrum, but slightly multimode at 405 nm. This did not significantly impact the experiments, and the optical crosstalk between the primary and adjacent fiber cores at 405 nm was \textless-12 dB (e.g., when the target grating has an emission power of 15 {\textmu}W, the emission power from an adjacent grating emitter is \textless0.76 {\textmu}W, which is insufficient for photoactivation and does not impact the localization of photoactivation). Overall, the localized and addressable photolysis demonstrated here shows the potential of photonic neural probes with microfluidic channels for localized photoactivation of caged dyes for fluorescence microscopy at a high spatial resolution, as well as target-specific release of caged drugs and neurotransmitters for neuronal activity manipulation \citep{martens2004caged,hess2009ultrahigh}.

\subsection{Uncaging in Fixed Brain Tissue}
\begin{figure}[ht]
\begin{center}
\includegraphics[width=1\textwidth]{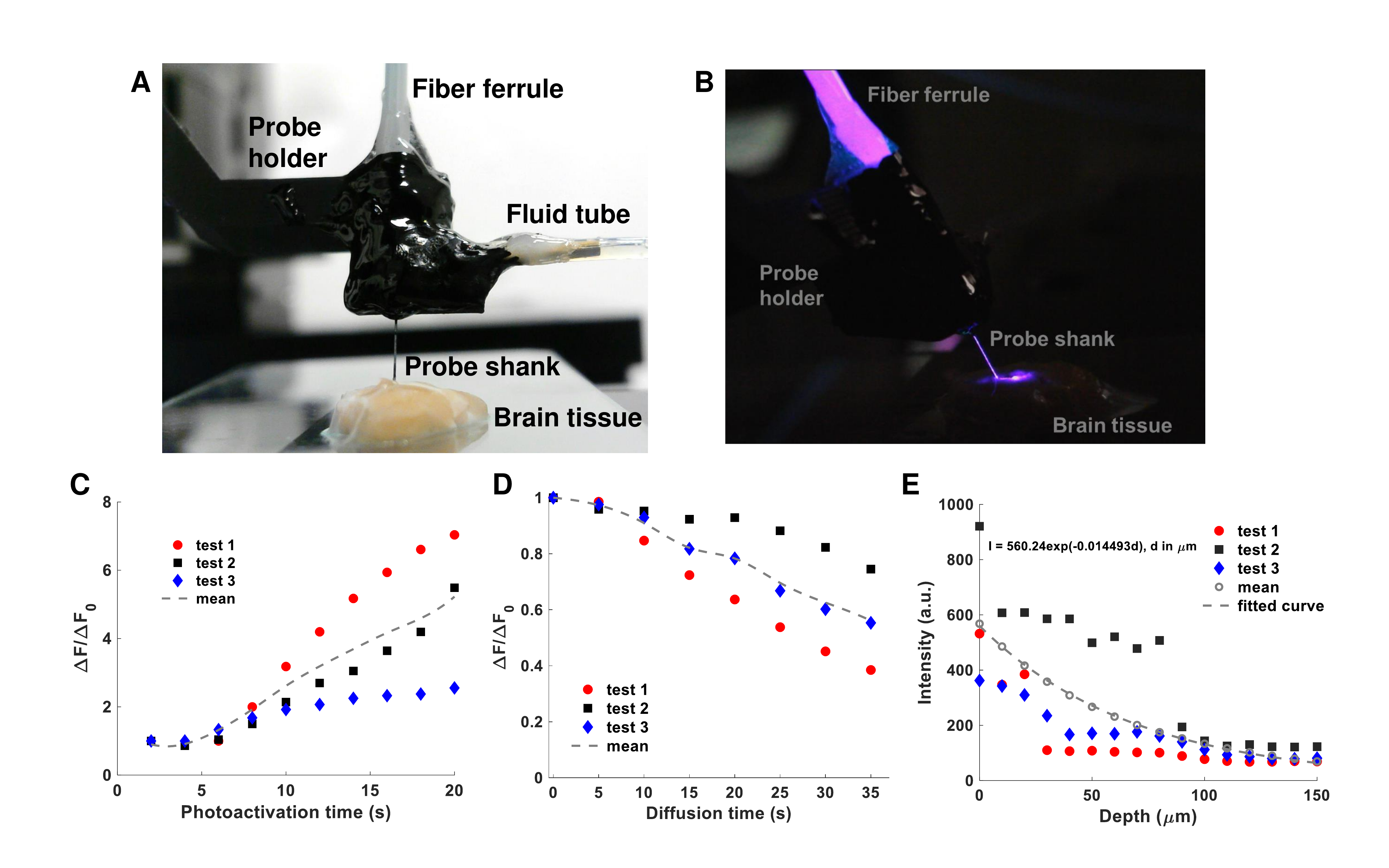}%
\end{center}
\caption{Uncaging with a neural probe (Probe 2) in fixed brain tissue. \textbf{(A)} Front-view photograph of the packaged neural probe inserted in a fixed brain slice. The CMNB-fluorescein solution was injected into the tissue via the integrated microfluidic channel. \textbf{(B)} Side-view photograph of 405-nm photoactivation light emitted from the neural probe in the brain tissue; the room light was turned off for increased visibility of the optical emission. \textbf{(C)} Normalized fluorescence intensity change \textit{$\Delta$F}/\textit{$\Delta$F$_0$} with sequential photoactivation (2 s pulses, 15 {\textmu}W grating emission into the sample); measured with a fluorescence microscope above the sample as in \textbf{Figures \ref{fig:5}}-\ref{fig:6}). \textit{$\Delta$F} is the mean pixel intensity in the selected ROI after background subtraction, and \textit{$\Delta$F$_0$} is the mean pixel intensity after the first photoactivation pulse. \textbf{(D)} Normalized fluorescence intensity change \textit{$\Delta$F/\textit{$\Delta$F$_0$}} during diffusion of the uncaged fluorescein in the brain tissue following the photoactivation pulses in \textbf{(C)}. The last data points of \textbf{(C)} correspond to the first of \textbf{(D)}, and no photoactivation light was applied in \textbf{(D)}. The dotted curves are the average of three experiments. \textbf{(E)} Attenuation of the optical intensity of a 405-nm beam emitted by the neural probe in the fixed brain slice I vs. propagation distance d. The optical intensities were extracted from micrographs of beam profiles with background subtraction. Data was collected from three insertion locations in the brain slice, and the dotted curve is the fitted exponential function following averaging of the three data sets.}
\label{fig:7}
\end{figure}

In our final uncaging experiment, the fluidic and photonic functionalities of the neural probe were used together for uncaging in a fixed brain slice from a wild-type mouse. The microfluidic channel injected CMNB-caged fluorescein and a 405-nm beam from one of the grating emitters performed photoactivation. The imaging setup in \textbf{Figure \ref{fig:5}A} was used for this uncaging experiment, and a front-view photograph of the neural probe inserted into the 2-mm thick fixed brain slice is shown in \textbf{Figure \ref{fig:7}A}. The brain slice preparation is described in Section \ref{brain slice preparation}, and the brain tissue was glued onto a clean glass slide with an adhesive (Insta-Cure+ Super Glue, Bob Smith Industries, Atascadero, CA, USA). Droplets of 1x PBS were applied on and around the fixed brain tissue to keep it moist during the experiments. \textbf{Figure \ref{fig:7}B} is a side-view photograph of the neural probe inserted into the brain slice with 405-nm optical emission from one of the grating emitters.

The neural probe shank was inserted into the fixed brain tissue via translation of the micromanipulator, and the implantation depth was monitored via the micromanipulator position readout. The grating emitter closest to the microfluidic channel outlet was used for this experiment. The implantation depth was selected so that the grating emitter was at a 50 {\textmu}m depth, a shallow depth where the outlet of the printed channel was also within the brain tissue for proper injection of the caged fluorescein solution. A shallow grating emitter depth avoided significant optical attenuation of the fluorescence from the uncaged fluorescein due to absorption and scattering in the fixed tissue. Prior to optical emission from the neural probe, a 100 {\textmu}M CMNB-caged fluorescein solution was injected into the brain tissue via the printed microfluidic channel for 120 s at a flow rate of 10 {\textmu}L/min (driven by the syringe pump). A 2.5 mL gastight glass syringe (1002 TLL, Hamilton Company) was installed on the syringe pump with a 24-gauge Luer Lock needle (549-0560, VWR, Avantar, Germany) to provide precise and stable fluid delivery. The syringe pump was also automated using Matlab R2021a. After the experiments, the neural probe was cleaned with 1\% w/v  Tergazyme solution (Z273287, Sigma-Aldrich) to remove the brain tissue adhered to the probe shank.
\label{uncage in brain}

\subsubsection{Sequential Photolysis of the Caged Fluorescein in Fixed Brain Tissue}
The photoactivation and fluorescence imaging workflow from Section \ref{sequential} was used in this experiment. In addition, the parameters for photoactivation and fluorescence imaging here were similar to Section \ref{sequential}: 15 {\textmu}W 405-nm grating emission power and 2-s duration for each photoactivation, 100 ms exposure time for the sCMOS camera, and the same epi-illumination power. The photoactivation was repeated 10 times during each test for a total photoactivation time duration of 20 s. As shown in \textbf{Figure \ref{fig:7}C}, the normalized fluorescence emission intensity change, \textit{$\Delta$F}/\textit{$\Delta$F$_0$}, progressively increased with the cumulative photoactivation time. The rate of increase in \textit{$\Delta$F}/\textit{$\Delta$F$_0$} was reduced after several photoactivation events, and saturation of \textit{$\Delta$F}/\textit{$\Delta$F$_0$} is expected as the 405-nm photoactivation fluence increases and the fraction of photolyzed caged fluorescein molecules in the illumination volume approaches 100\% \citep{trigo2009laser}. Similar to Section \ref{sequential}, \textit{$\Delta$F} is the mean pixel intensity in the selected ROI containing the uncaged fluorescein profiles, and the definition of the ROI is shown in Supplementary \textbf{Figure S4}. \textit{$\Delta$F$_0$} is the mean pixel intensity in the ROI after the first photoactivation pulse. Three tests were performed in the brain slice, and for each test, the neural probe was inserted at a different site free of uncaged fluorescein from previous tests. We hypothesize that the observed variation in \textit{$\Delta$F}/\textit{$\Delta$F$_0$} in the three repeated tests is due to anatomical differences between the insertion sites and corresponding differences in the optical attenuation of the emitted beam. In addition to the results presented in \textbf{Figure \ref{fig:7}C}, tests were repeated in 4 other brain slices, and fluorescence intensity increases during photoactivation were also observed in these slices. Compared with photolysis of caged fluorescein in solution (Supplementary \textbf{Figure S3}), photolysis of injected caged fluorescein in fixed brain tissue (Supplementary \textbf{Figure S4}) resulted in lower fluorescence intensity changes in the field of view, which can be attributed to the smaller number of caged fluorescein molecules in the small volume of injected solution and the high optical attenuation in the brain tissue. The optical attenuation reduces both the photoactivation intensity in the brain tissue and the amount of fluorescence collected by the microscope.

\subsubsection{Characterization of Uncaged Fluorescein Diffusion in Brain Tissue}
Similar to Section \ref{diffusion}, the diffusion of the uncaged fluorescein in brain tissue was studied. For each test in the previous section, following 20 s of cumulative photoactivation of the injected caged fluorescein, the grating emitter was turned off, and a fluorescence image sequence was captured with a 5-s time interval between images (with the microscope epi-illumination turned on briefly for each image). The normalized fluorescence intensity change \textit{$\Delta$F/\textit{$\Delta$F$_0$}} during the diffusion process is shown in \textbf{Figure \ref{fig:7}D}. Compared with the diffusion of uncaged fluorescein in solution (\textbf{Figure \ref{fig:5}F}), the released free-form fluorescein molecules diffused remarkably slower in the fixed brain slice. A possible mechanism for the variation of fluorescence intensity change, and hence diffusion, in the three repeated tests is the diffusion anisotropy in the fixed brain tissue \citep{sun2005formalin}.

\subsubsection{Characterization of Optical Power Attenuation in the Fixed Brain Tissue Sample}
To characterize the optical attenuation caused by absorption and scattering in the fixed brain tissue samples used here, scattered emission light from the probe was measured with increasing depth of the grating emitter in the tissue. The piece of fixed brain tissue used in these measurements had not received prior caged fluorescein injections. The neural probe was implanted into the brain tissue via the micromanipulator with a depth increment of 10 {\textmu}m. One of the grating emitters was turned on and emitted a 405-nm beam into the brain tissue. At each depth step, an image was acquired using the top-down wide-field microscope; the filter cube was removed for imaging of the scattered 405 nm light from the brain tissue sample. The measured scattered light intensity (with background subtraction) versus depth is shown in \textbf{Figure \ref{fig:7}E}. The measurements were repeated three times at different implantation sites in the brain tissue sample. Each intensity vs. depth data set was fitted by an exponential curve, and the three exponential curves were averaged. The fitted power attenuation coefficient corresponds to an optical attenuation length of $\sim$69 {\textmu}m. The optical attenuation length is a function of multiple parameters, including the optical wavelength, brain region, and the preparation method of the brain tissue \citep{al2013light,yona2016realistic}.
\label{power decay}

\section{Discussion}

\subsection{Emission Power Variation at 405 nm}
During the uncaging experiments, we observed a time-dependent drift in the emitted optical power of the grating emitters. The effect was observed at a wavelength of 405 nm and not at 473 nm. Generally, the emission power of the grating emitters dropped with increasing emission time at 405 nm. The rate of power reduction varied across the grating emitters, and emitters with higher transmission experienced a slower power reduction. For instance, a grating emitter with a transmission of $\num{-25}$ dB on Probe 2 maintained a stable emission power of $\sim$9.1 {\textmu}W for 180 s, while another grating emitter on the same probe with a transmission of $\num{-32}$ dB and an initial emission power of $\sim$5.8 {\textmu}W, experienced a $\sim$90\% power drop within 90 s (Supplementary \textbf{Figure S5}).

We eliminated photodarkening \citep{koponen2006measuring} of the on-chip SiN waveguides as a possible mechanism by aligning a single-mode fiber (S405-XP, Thorlabs) to an edge coupler of a photonic neural probe chip and injecting 405-nm laser light corresponding to $\sim$20.5 {\textmu}W grating emission power (higher than the emission power used during uncaging experiments). After 30 min of emission, no reduction in the emitted optical power was observed. Photodarkening of the MCF was eliminated next. 405-nm laser light was coupled to a core of the MCF with a high output power at the distal facet of $\sim$14 mW, and the output power remained stable for a period of 30 minutes (Supplementary \textbf{Figure S5}).

We hypothesize that the optical emission power reduction was due to degradation of the epoxy used for fiber attachment and probe encapsulation under exposure to 405-nm light. A significant fraction of the 405-nm light (near-UV) from the MCF was not coupled into the waveguides at the fiber-to-probe interface and was scattered inside the epoxy encapsulation, possibly further hardening or burning the UV-curable epoxy and causing continuous shrinkage or deformation of the epoxy. Subtle degradation of the epoxy may result in misalignment between the fiber cores and the edge couplers, resulting in lower output powers. 

In the proof-of-concept demonstrations in this work, the packaging protocol and selection of epoxies was based on our previous photonic neural probe work \citep{sacher2021implantable,sacher2022optical,chen2022implantable} for optogenetic stimulation and light sheet imaging (473-488 nm wavelengths). Modifications to the packaging strategy and epoxy selection for increased tolerance to 405-nm light may reduce the magnitude of emission power variations, potentially leading to more robust and long-lasting neural probes for uncaging.

\subsection{Spatial Distribution of Microfluidic Channel Outlet and Grating Emitters}
To prevent coverage of grating emitters by the 3D-printed microfluidic channel, the channel outlet was $\sim$1.5 mm away from the tip of the neural probe shank for Probe 2 (Supplementary \textbf{Figure S1}), ensuring unobstructed optical emission from all 16 grating emitters. In the uncaging characterization method in this work, fluorescence imaging of uncaged fluorescein was performed with a top-down microscope. Limited by epifluorescence imaging depth (due to tissue scattering), only uncaged fluorescein in close proximity to the tissue surface (and channel outlet) could be properly imaged. Considering the 110-{\textmu}m longitudinal pitch of grating emitters on Probe 2, this limited our uncaging characterization method in fixed brain tissue to, at most, 2 grating emitters. This is a limitation of the characterization method and its dependence on fluorescence imaging, not a limitation of uncaging with photonic neural probes and integrated microfluidic channels. Since (with sufficient time) the diffusion profile of injected caged compounds from the microfluidic channel can overlap all grating emitters in the 1.5-mm distance spanning the tip of the shank to the channel outlet, all 16 grating emitters may, in principle, be used for uncaging.

\subsection{Potential Applications of Uncaging with Photonic-Microfluidic Neural Probes}
In this work, we demonstrated photolysis of caged fluorescein at 405 nm as a proof of concept of uncaging with photonic-microfluidic neural probes. Photoactivation of a caged dye enables simple quantification of the uncaging via fluorescence imaging and is well-suited to an initial demonstration. Applications for highly-localized fluorescence microscopy via injection of caged dyes in tissue are possible, but are not the primary motivation for this work. Our future development of neural probes for uncaging is directed toward a variety of novel local uncaging experiments of caged drugs (antagonists/agonists) and neurotransmitters (e.g., glutamate, dopamine) to further investigate local manipulations on neural microcircuits.

\begin{figure}[ht]
\begin{center}
\includegraphics[width=1\textwidth]{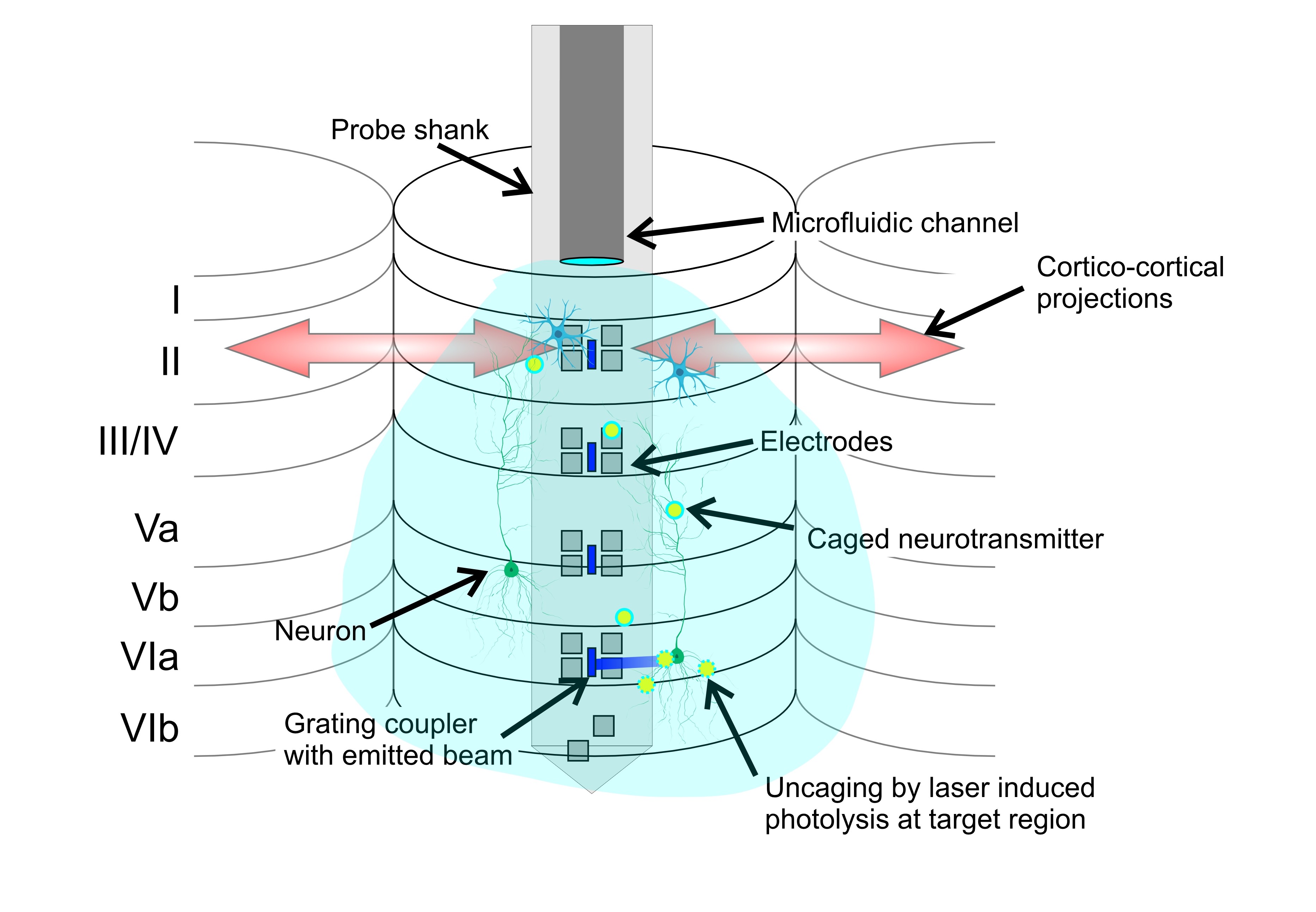}%
\end{center}
\caption{Conceptual illustration of the potential application of the neural probe to cortical microcircuits investigation by local photolysis of caged neurotransmitters (not to scale). The cortical layers \Romannum{1} to \Romannum{6}, cortical columns, and cortico-cortical projections are shown. Grating emitters and electrodes span the cortical layers, and the microfluidic channel injects a caged compound, which diffuses throughout the cortical column. The addressable laser beams emitted by the probe enable precise and targeted photolysis of the caged compound.}
\label{fig:8}
\end{figure}

Uncaging with photonic neural probes can be particularly relevant to the investigation of cortical microcircuits on the columnal level (\textbf{Figure \ref{fig:8}}). Current neurochemical approaches consist primarily of topical applications of agonists and antagonists resulting in a global affection of nearby cortical columns, eventually affecting the cortico-cortical onsite integration at the recording site. A primary limitation is that the diffusion of drugs through the cortical layers may result in minutes of delay until lower regions are reached, impacting the upper regions during the diffusion process \citep{happel2014dopamine, deane2020ketamine}. Other more localized approaches apply the drugs directly via injection onsite the recording electrodes, requiring a high degree of accuracy in the placement of the injection site to avoid affecting the non-target regions. Uncaging with photonic neural probes avoids these limitations. The caged compound is injected in close proximity to the light emitters and electrodes, and the addressable laser beams emitted by the neural probe perform local uncaging in the target regions/layers, without affecting the non-target regions/upper layers. 

In the remainder of this subsection, we discuss an example of neurotransmitter uncaging in relationship to our demonstrated neural probe performance. Caged neurotransmitters are important to manipulate intracellular biochemistry and mimic intercellular signaling in living tissue \citep{lutz2008holographic}. Among the various photosensitive neurotransmitters, 4-methoxy-7-nitroindolinyl (MNI) caged glutamate has been commonly used to record photolysis-evoked currents. Photolysis of MNI-caged glutamate at 405 nm has been reported with an energy density of 20 J/cm\textsuperscript{2} (0.1 ms pulse duration, 2 mW/\textrm{\textmu}m\textsuperscript{2} intensity, \cite{trigo2009laser}) and 10 J/cm\textsuperscript{2} \citep{lutz2008holographic}. In our uncaging experiment in fixed brain tissue, each pulse corresponded to an energy density of 20 J/cm\textsuperscript{2} (2 s pulse, 15 {\textmu}W emission power, 150 \textrm{\textmu}m\textsuperscript{2} emission area) before correcting for the optical attenuation in brain tissue. The estimated energy density of a 2-s pulse is 9.1 J/cm\textsuperscript{2} after a 50 {\textmu}m propagation distance in brain tissue (power attenuation coefficient \textit{\textmu}=0.0145/{\textmu}m, \textbf{Figure \ref{fig:7}E}). Thus, the photonic-microfluidic neural probes in this work show promise for the photolysis of caged glutamate, with comparable energy densities to previous reports, albeit with larger pulse durations. Optimization of fiber-to-chip coupling efficiencies and waveguide losses at 405 nm may result in a significant increase in optical transmission and corresponding reduction in the required pulse durations. Moreover, additional work towards mounting the probe chip on a printed circuit board carrier and wire bonding the on-chip electrical pads will enable access to the microelectrodes on the probe shank and the functionality of extracellular recording of photolysis-evoked neuronal activity.

\subsection{Phototoxicity of 405-nm Photolysis}
Phototoxicity is a critical limiting factor for laser photolysis with high spatial resolution and short wavelengths, leading to photodamage effects in live-cell experiments \citep{mohr2016labeling}. In \cite{trigo2009laser}, an energy dose of 800 J/cm\textsuperscript{2} (0.1 ms pulse, 2 mW/\textrm{\textmu}m\textsuperscript{2} intensity, 0.2 Hz, 200 s duration) for complete photolysis of MNI-caged glutamate did not show evidence of toxicity to cerebellar molecular layer interneurons or Purkinje neurons. In our uncaging experiment with fixed brain tissue, each photoactivation pulse delivered an estimated energy density of 9.1 J/cm\textsuperscript{2} following 50 {\textmu}m of propagation in brain tissue. Proper control over the number of photoactivation pulses at the same photolysis site may enable reliable photolysis with neural probes in living brain tissue with non-toxic optical energy doses.

\section{Conclusion}

In summary, we have proposed and demonstrated the integration of microfluidic channels onto foundry-fabricated photonic neural probes via two photon polymerization 3D printing. This strategy enables the integration of microfluidics functionalities onto state of the art foundry-fabricated photonic and electrophysiological neural probes without requiring expensive and time-consuming modifications to foundry fabrication process flows. Through a series of validation experiments, we have shown and characterized the optical and fluidic functionalities of our photonic neural probes with integrated microfluidic channels. Addressable, low-divergence beams emitted from the grating emitters on the probes are well-suited to target-specific optogenetic stimulation and/or uncaging, while the microfluidic channels deliver well-defined diffusion profiles of injected chemicals in close proximity to the grating emitters with reasonable flow resistances, providing a convenient and precise method for injection of drugs and caged compounds. As a proof-of-concept demonstration, the probes were validated for 405-nm photolysis of caged fluorescein in solution and in fixed brain tissue, showing localized and addressable uncaging. Overall, the demonstrated neural probes provide a unique combination of high target specificity at depth and minimal tissue displacement compared to conventional methods of light and fluid delivery - presenting new possibilities for neuroscience experiments employing photolysis and/or optogenetic stimulation. Potential future use cases discussed in this work include: 1) photolysis of caged glutamate at non-toxic optical intensities for the investigation of activation kinetics of glutamate receptors \textit{in situ} and the role of individual neurons in neural circuits, and 2) the local manipulation of cortical columns via drug induced changes in the cortical processing of cortical microcircuits. The target specificity of the probe-enabled uncaging allows the investigation of local effects on cortical columns while keeping the cortico-cortical interactions intact. Future technology development work aims to demonstrate simultaneous photonic, electrophysiological, and microfluidic functionalities with our neural probes, in addition to multi-wavelength optical emission, broadening the range of functionalities and compatible neuroscience experiments.

\section*{Conflict of Interest Statement}
The authors declare that the research was conducted in the absence of any commercial or financial relationships that could be construed as a potential conflict of interest.

\section*{Data Availability Statement}
The original contributions presented in the study are included in the article/supplementary material, further inquiries can be directed to the corresponding authors.

\section*{Ethics Statement}
Brain tissues have been kindly provided by the Fraunhofer Institute for Cell Therapy and Immunology IZI and were handled in accordance with the German laws.

\section*{Author Contributions}

WS conceived the project. XM and KD developed the 3D printing workflow. FC, XM, and WS designed the neural probes. XL, HC, and GL fabricated the neural probes. XM, FC, and PD developed the probe assembly approach. XM, JP, and WS designed the experiments. XM, MB, JL, HW, and AS built the experimental setup. XM performed the experiments and analyzed the results. XM, MB, JP, and WS wrote the manuscript with inputs from other co-authors. JP and WS supervised the project. All authors contributed to the article and approved the submitted version.

\section*{Funding}
This work was supported by the Max Planck Society.

\section*{Acknowledgments}
The authors thank Dr. Holger Cynis, Dr. Ines Koska and Dr. Stefanie Gei{\ss}ler from the Fraunhofer Institute for Cell Therapy and Immunology IZI for providing the brain tissue samples from mice. The authors also acknowledge helpful discussions and technical support from Alperen G{\"o}vdeli, Ankita Sharma, Tianyuan Xue, and Frank Weiss at the Max Planck Institute of Microstructure Physics.

\bibliographystyle{Frontiers-Harvard} 
\bibliography{reference}

\section*{Figure captions}

\clearpage

\section*{Supplementary Material}
The Supplementary Material for this article can be found online at \url{https://doi.org/10.17617/3.5IOMQM}
\label{supplementary}

\subsection*{Supplementary Data}

CAD file. Design file of the microfluidic channel with a channel inlet located on the neural probe base and a channel outlet on the shank (Probe 1). The unit in the .stl file is mm.

Video 1. Front-view video of the diffusion of a red dye injected into an agarose gel block by a photonic neural probe with a microfluidic channel (Probe 1). Probe 1 was inserted into the 1\% w/v agarose gel block to a depth of $\sim$1.5 mm. The outlet of the microfluidic channel was inside the agarose gel. The 4 mM Allura red dye solution was injected at a flow rate of 10 {\textmu}L/min using a syringe pump. The video is real-time.

Video 2. Top-down microscope video of simultaneous light emission and fluid delivery with Probe 3. Probe 3 was fabricated on the same wafer as Probe 1, with a nominal SiN thickness of 200 nm. Probe 3 had 2 columns of grating emitters (16 emitters in total, grating pitch 440 nm, staggered on 55 {\textmu}m lateral and longitudinal inter-grating pitches). During the video, Probe 3 was mounted horizontally with a 473-nm laser input to the probe. Deionized water was manually injected through the microfluidic channel while switching between grating coupler emitters. The video was compiled from image sequences and is played at 2x speed.

\subsection*{Supplementary Tables and Figures}

\setcounter{figure}{0}

\begin{figure}[ht]
\renewcommand{\thefigure}{S\arabic{figure}}
\begin{center}
\includegraphics[width=1\textwidth]{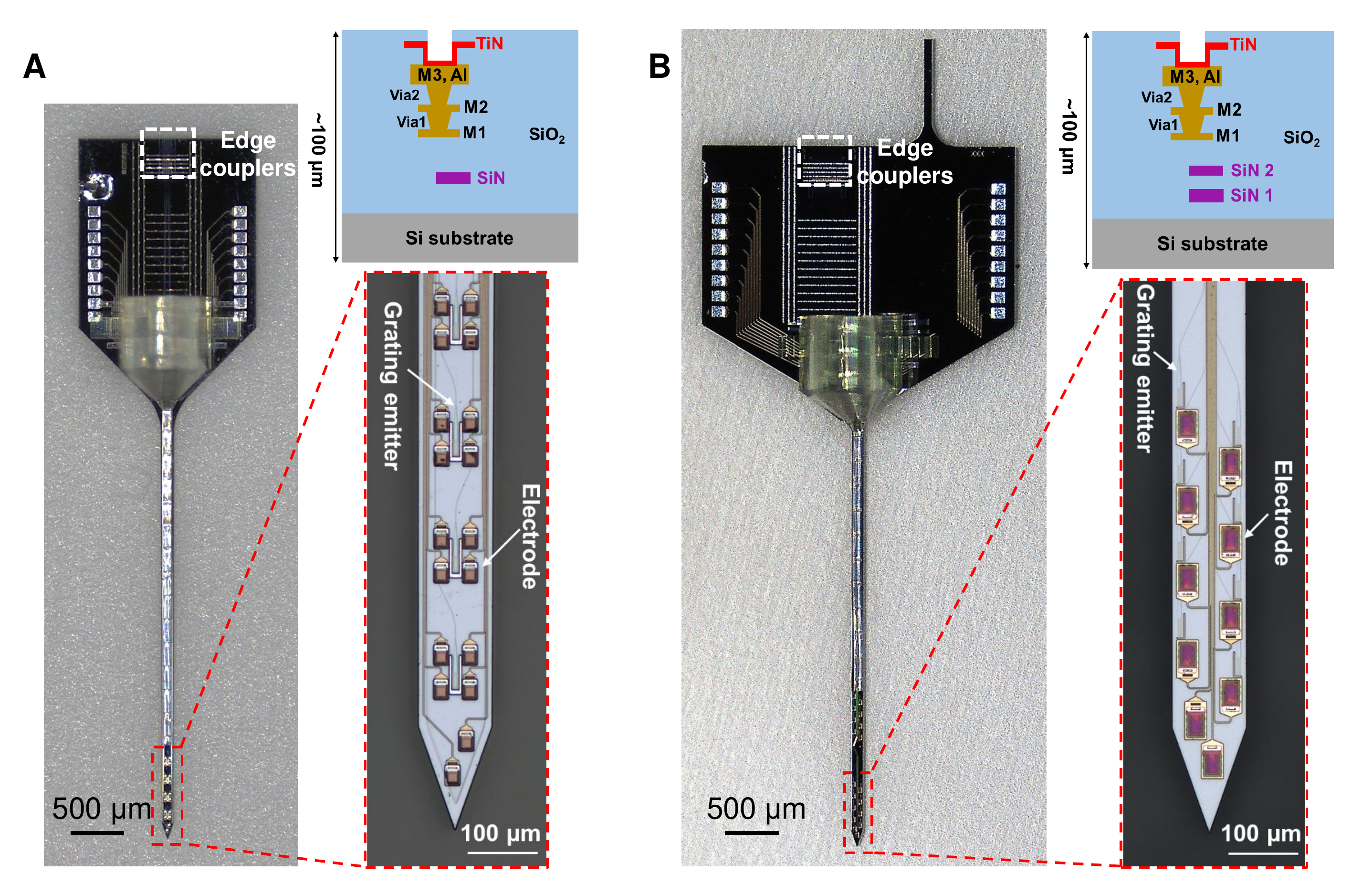}%
\end{center}
\caption{Additional details of the neural probe designs. Comparison of \textbf{(A)} Probe 1 and \textbf{(B)} Probe 2 with optical micrographs and illustrations of the probe cross-sections. (Insets) Optical micrographs of the tips of Probe 1 and Probe 2 shanks showing grating coupler emitters and electrodes.}
\label{fig:S1}
\end{figure}

\begin{figure}[ht]
\renewcommand{\thefigure}{S\arabic{figure}}
\begin{center}
\includegraphics[width=1\textwidth]{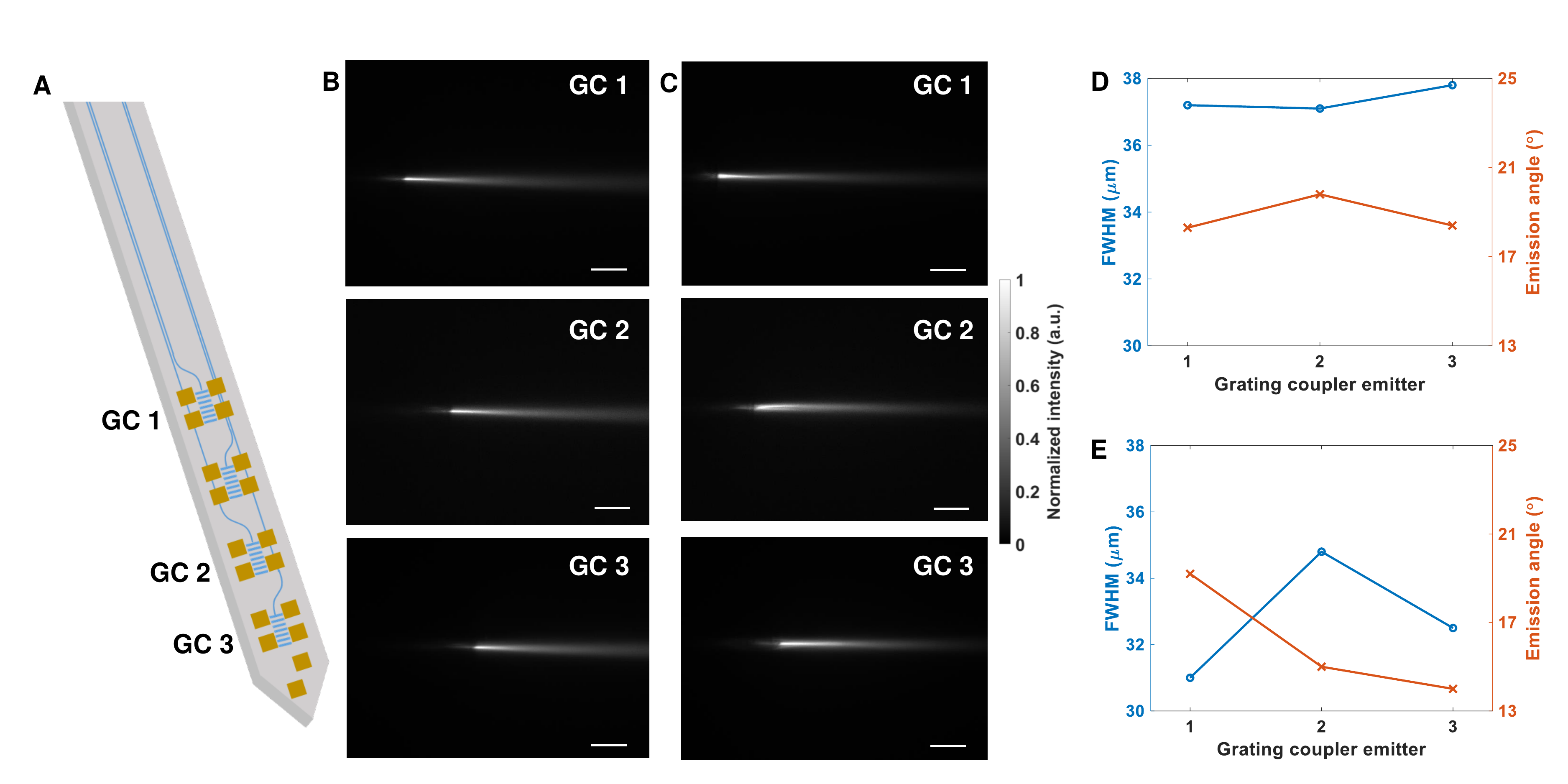}%
\end{center}
\caption{Additional details of the measured beam profiles of Probe 1. \textbf{(A)} Illustration of the shank tip of Probe 1 (not to scale). Top-down fluorescence images of beam profiles from Probe 1 in \textbf{(B)} a fluorescein solution and \textbf{(C)} agarose gel. The scale bars are 100 {\textmu}m. Beam profiles from grating coupler 1 (GC1), GC2, and GC3 are shown (the term ``grating coupler" is synonymous with ``grating emitter," which is used elsewhere in the manuscript). Drops of fluorescein were placed on the top surface of the agarose gel and diffused into the gel, enabling fluorescence imaging of the beam profiles. Full width at half maximum (FWHM) after a 300 {\textmu}m propagation distance and emission angle measurements for Probe 1 emitted beams in \textbf{(D)} fluorescein and \textbf{(E)} agarose gel. The measured FWHMs and emission angles in agarose gel have a larger variation, possibly due to impurities and structural inhomogeneity in the agarose gel block.}
\label{fig:S2}
\end{figure}

\begin{figure}[ht]
\renewcommand{\thefigure}{S\arabic{figure}}
\begin{center}
\includegraphics[width=1\textwidth]{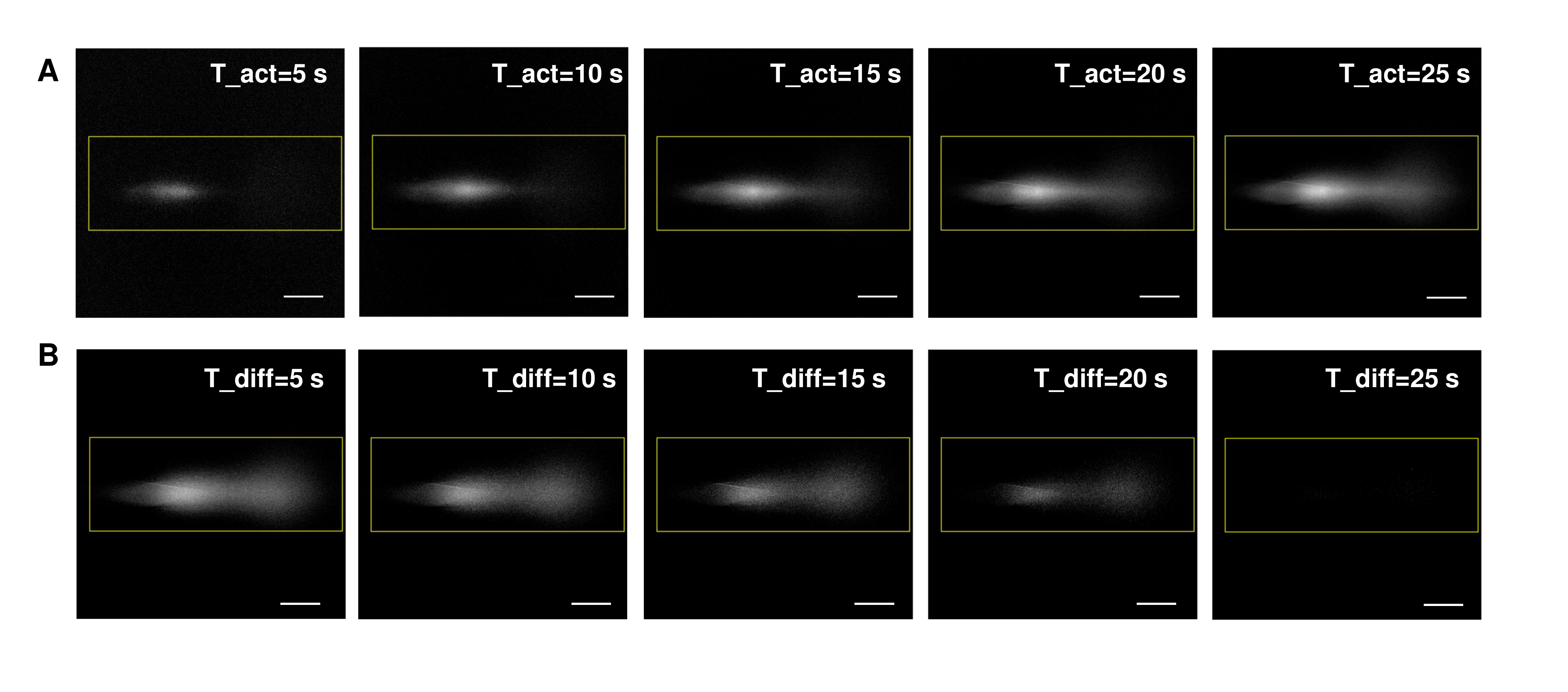}%
\end{center}
\caption{Additional details of fluorescence intensity change (\textit{$\Delta$F}) analysis for uncaging in a CMNB-caged fluorescein solution bath (\textbf{Figure 5}). \textbf{(A)} and \textbf{(B)} show time-dependent fluorescence profiles with background subtraction after photoactivation and during diffusion in test 1, respectively. The scale bars are 200 {\textmu}m. The region of interest (ROI) for calculation of \textit{$\Delta$F} is delineated by the yellow rectangle and was selected to avoid cropping the fluorescence profile during sequential photoactivation. \textit{$\Delta$F} was calculated as the mean pixel intensity in the ROI. Sequential micrographs of the fluorescence profiles are shown after \textbf{(A)} accumulated photoactivation time (T\textunderscore act) and \textbf{(B)} diffusion time (T\textunderscore diff) of 5 s, 10 s, 15 s, 20 s, and 25 s. The same ROI was applied for \textit{$\Delta$F} analysis of the three tests in \textbf{Figure 5}. The raw images are available at \url{https://doi.org/10.17617/3.5IOMQM}.}
\label{fig:S3}
\end{figure}

\begin{figure}[ht]
\renewcommand{\thefigure}{S\arabic{figure}}
\begin{center}
\includegraphics[width=1\textwidth]{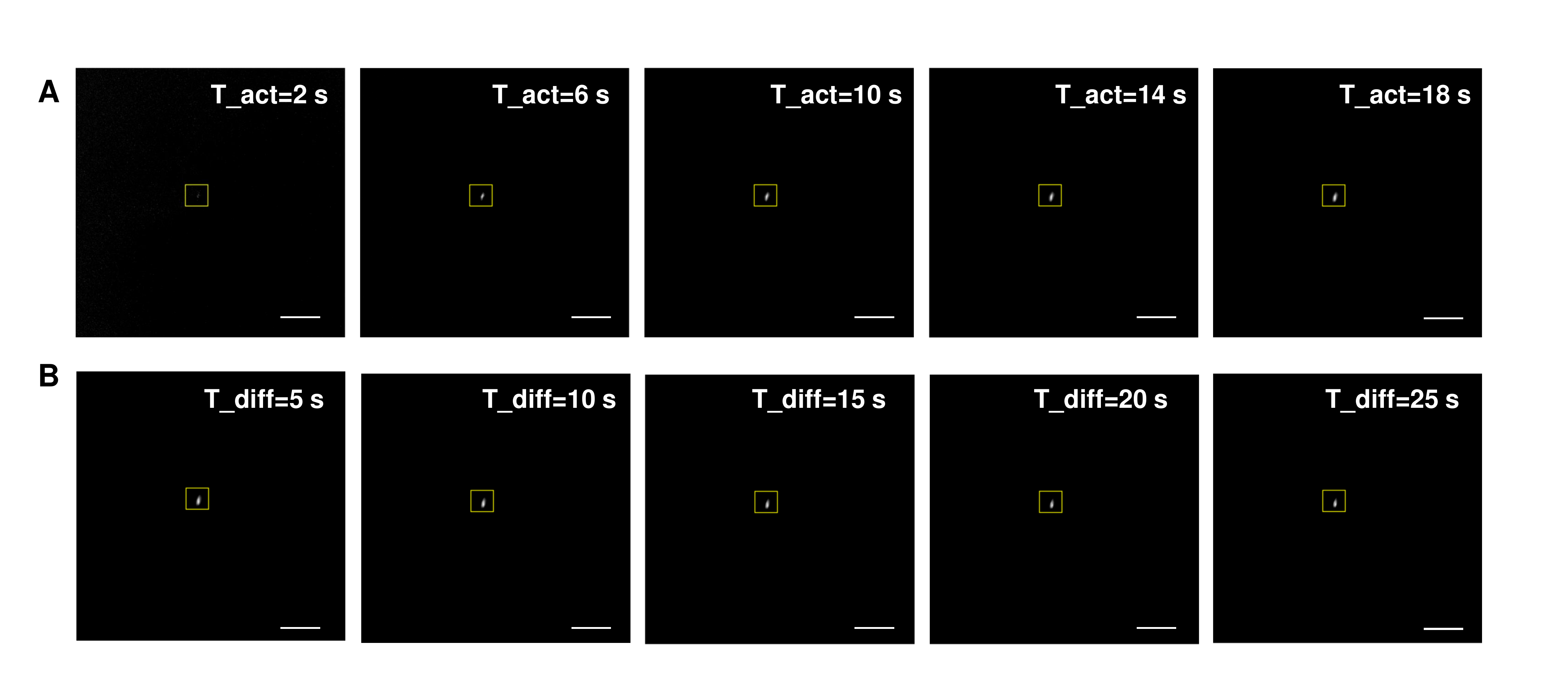}%
\end{center}
\caption{Additional details of fluorescence intensity change (\textit{$\Delta$F}) analysis for uncaging in fixed brain tissue (\textbf{Figure 7}). \textbf{(A)} and \textbf{(B)} show time-dependent fluorescence profiles with background subtraction after photoactivation and during diffusion in test 1, respectively. The scale bars are 200 {\textmu}m. The ROI was selected to include the fluorescence profile during sequential photoactivation. Sequential micrographs of the fluorescence profiles are shown with time stamps labeled accordingly. The same ROI was applied for \textit{$\Delta$F} analysis of the three tests in \textbf{Figure 7}. The raw images are available at \url{https://doi.org/10.17617/3.5IOMQM}.}
\label{fig:S4}
\end{figure}

\begin{figure}[ht]
\renewcommand{\thefigure}{S\arabic{figure}}
\begin{center}
\includegraphics[width=1\textwidth]{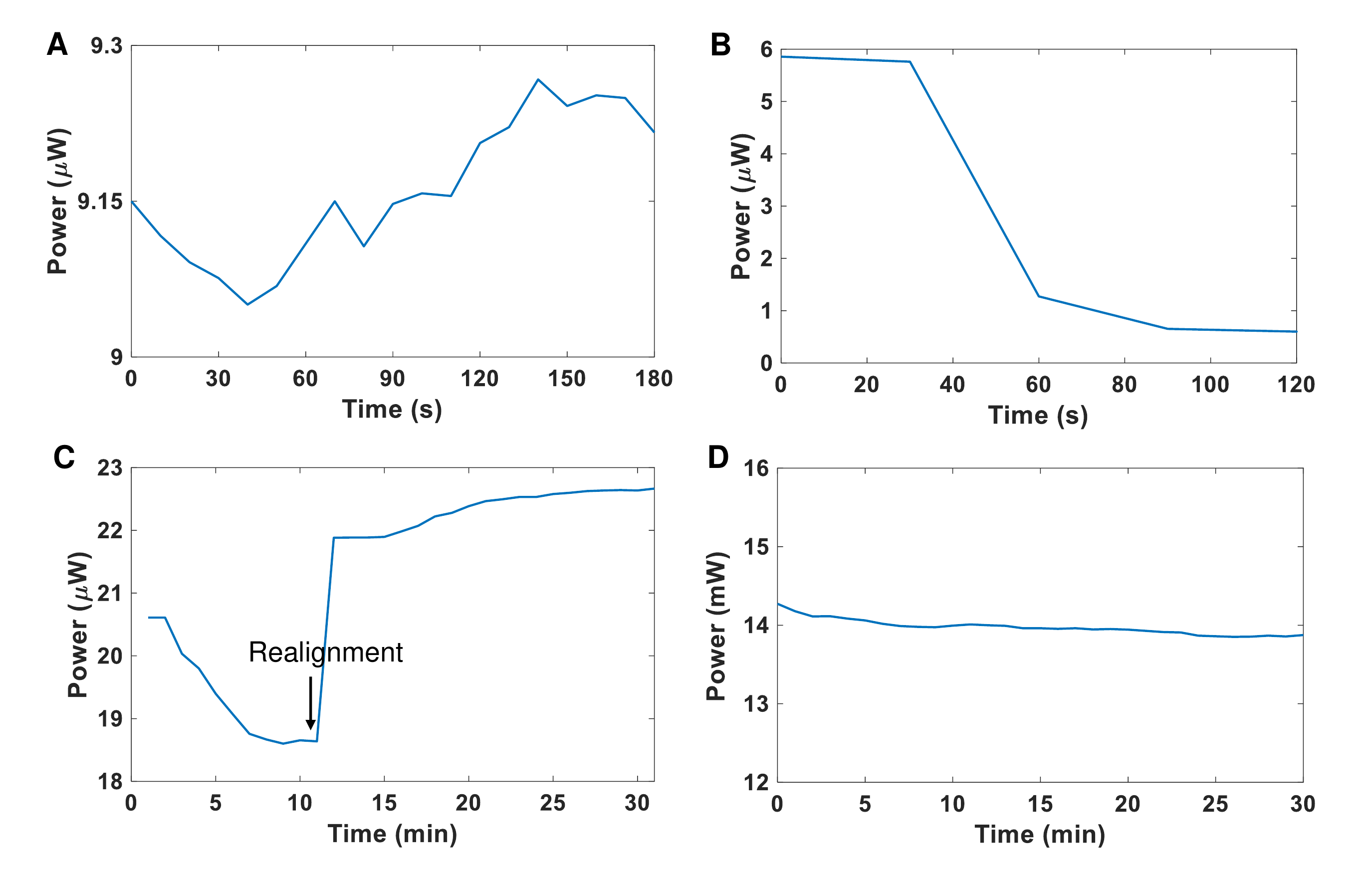}%
\end{center}
\caption{Optical power stability test at 405 nm. Emission power of a Probe 2 grating coupler emitter corresponding to \textbf{(A)} $\num{-25}$ dB and \textbf{(B)} $\num{-32}$ dB optical transmission (with transmission defined as the ratio of emitted power from the grating coupler and input power to the optical scanning system). \textbf{(C)} Emission power of a grating coupler emitter from a neural probe before packaging. The neural probe under test was extracted from the same wafer as Probe 2, and 405-nm laser light was coupled to the probe chip from a single mode fiber. The alignment was re-adjusted at t=11 min. \textbf{(D)} Output power of a core of the multicore optical fiber, without attachment to a neural probe.}
\label{fig:S5}
\end{figure}

\end{document}